\documentclass[pra,twocolumn,reprint,showpacs,superscriptaddress,aps]{revtex4-1}

\usepackage[applemac]{inputenc}
\usepackage[T1]{fontenc}
\usepackage{color}
\usepackage{amsmath}
\usepackage{amssymb}
\usepackage{graphicx}
\usepackage{float}
\usepackage{bm}

\begin{document}

\newcommand{\bra}[1]{\left\langle{#1}\right|}
\newcommand{\ket}[1]{\left|{#1}\right\rangle}
\newcommand{\id}{\openone}
\newcommand{\eq}[1]{(\ref{#1})}
\newcommand{\vphi}{\varphi}
\newcommand{\hc}{\text{H.c.}}
\newcommand{\bo}{\boldsymbol{(}}
\newcommand{\bc}{\boldsymbol{)}}

\newcommand{\ad}{a^\dagger}
\newcommand{\sz}[1]{\sigma_{z}^{#1}}
\newcommand{\sx}[1]{\sigma_{x}^{#1}}
\newcommand{\sy}[1]{\sigma_{y}^{#1}}
\newcommand{\sign}[1]{\mathrm{sgn}\left( #1 \right)}
\newcommand{\sinInt}[1]{\mathrm{Si}\left( #1 \right)}
\newcommand{\cosInt}[1]{\mathrm{Ci}\left( #1 \right)}
\newcommand{\etal}{{\it et. al}}
\newcommand{\tr}[1]{\text{Tr}\left[ {#1} \right]}
\newcommand{\average}[1]{\left\langle {#1} \right\rangle}
\newcommand{\kevmax}[2]{\text{Max}_{#1}\left[ {#2} \right]}
\newcommand{\trs}[2]{\text{Tr}_{#2}\left[ {#1} \right]}

\newcommand{\red}{\color[rgb]{0.8,0,0}}
\newcommand{\green}{\color[rgb]{0.0,0.6,0.0}}
\newcommand{\blue}{\color[rgb]{0.0,0.0,0.6}}
\newcommand{\greenBlue}{\color[rgb]{0.,0.5,0.5}}
\newcommand{\redBlue}{\color[rgb]{0.5,0.5,0}}
\newcommand{\ab}[1]{{\red AB:~#1}} % AB's comments
\newcommand{\kl}[1]{{\green KL:~#1}} % KL's comments
\newcommand{\av}[1]{{\blue AV:~#1}} % AV's comments
\newcommand{\aw}[1]{{\greenBlue AW:~#1}} % AW's comments
\newcommand{\bs}[1]{{\redBlue BS:~#1}} % BS's comments
\newcommand{\nnn}{\nonumber \\}

\def\be{%
\begin{equation}}
\def\aout{a_{\text{out}}}
\def\ain{a_{\text{in}}}
\def\gammas{\gamma_{\mathrm{s}}}
\def\gammajs{\gamma^{j}_{\mathrm{s}}}
\def\gnr{\gamma_{\mathrm{nr}}}
\def\gr{\gamma_{\mathrm{r}}}
\def\gnrj{\gamma^{j}_{\mathrm{nr}}}
\def\ee{\end{equation}}
\def\tGamma{\tilde{\Gamma}}
\newcommand{\braket}[2]{\left\langle #1 | #2 \right\rangle}
\newcommand{\ketbra}[2]{\left| #1 \rangle \langle #2 \right|}
\newcommand{\mean}[1]{ \left\langle #1 \right\rangle}
\newcommand{\re}[1]{\text{Re}\left[ #1 \right]}
\renewcommand{\arg}[1]{\text{Arg}\left[ #1 \right]}
\newcommand{\im}[1]{\text{Im}\left[ #1 \right]}

\title{Input-output theory for waveguide QED with an ensemble of inhomogeneous atoms}

\date{\today}

\author{Kevin Lalumi\`ere}
\affiliation{D\'epartement de Physique, Universit\'e de Sherbrooke, Sherbrooke, Qu\'ebec, Canada J1K 2R1}
\author{Barry C. Sanders}
\affiliation{Institute for Quantum Science and Technology, University of Calgary, Alberta, Canada T2N 1N4}
\author{A. F. van Loo}
\affiliation{Department of Physics, ETH Zurich, CH-8093 Z\"urich, Switzerland}
\author{A. Fedorov}
\affiliation{Department of Physics, ETH Zurich, CH-8093 Z\"urich, Switzerland}
\author{A. Wallraff}
\affiliation{Department of Physics, ETH Zurich, CH-8093 Z\"urich, Switzerland}
\author{A. Blais}
\affiliation{D\'epartement de Physique, Universit\'e de Sherbrooke, Sherbrooke, Qu\'ebec, J1K 2R1, Canada}

\begin{abstract}
We study the collective effects that emerge in waveguide quantum electrodynamics where several (artificial) atoms are coupled to a one-dimensional superconducting transmission line. Since single microwave photons can travel without loss for a long distance along the line, real and virtual photons emitted by one atom can be reabsorbed or scattered by a second atom. Depending on the distance between the atoms, this collective effect can lead to super- and subradiance or to a coherent exchange-type interaction between the atoms. Changing the artificial atoms transition frequencies, something which can be easily done with superconducting qubits (two levels artificial atoms), is equivalent to changing the atom-atom separation and thereby opens the possibility to study the characteristics of these collective effects. To study this waveguide quantum electrodynamics system, we extend previous work and present an effective master equation valid for an  ensemble of inhomogeneous atoms driven by a coherent state. Using input-output theory, we compute analytically and numerically the elastic and inelastic scattering and show how these quantities reveal information about collective effects. These theoretical results are compatible with recent experimental results using transmon qubits coupled to a superconducting one-dimensional transmission line [A.~F.~van Loo {\it et al.}].
\end{abstract}

\pacs{42.50.Nn, 42.50.Lc, 71.70.Gm, 84.40.Az}

\maketitle

\section{Introduction}

By confining the electromagnetic field in space, cavity quantum electrodynamics (QED), and more recently circuit QED, have opened the opportunity to study the interaction of light and matter in the strong-coupling regime where the light-matter interaction strength overwhelms decay rates~\cite{haroche:2006a,wallraff:2004a}. Strong interaction between matter and propagating photons is of interest for applications such as quantum networks~\cite{kimble:2008a,tey:2008a,obrien:2009a} and single-photon transistors~\cite{chang:2007b,hwang:2009a}. Strong light-matter interaction in an open three-dimensional (3D) setting is made possible by tightly focusing the optical field~\cite{schuller:2010a}. An important signature of the interaction in this situation is the extinction of the transmitted light field by a single atom or molecule. Indeed, the light beam interferes destructively with the co-linearly emitted light from the atom or molecule, resulting ideally in 100\% reflection. However, because of poor spatial mode-matching (i.e.~the atom or molecule emits light in all directions while the incoming beam is tightly focused), only about 10\% reflection is currently observed with single atoms~\cite{tey:2008a}.

The situation can be very different with artificial atoms in a circuit~\cite{shen:2005a}. Indeed, as first shown experimentally by Astafiev \emph{et al}.~\cite{Astafiev:2010}, almost ideal mode matching can be realized with a superconducting flux qubit coupled to a one-dimensional (1D) transmission line. In that experiment, 94\% extinction of the transmitted signal was observed showing that a single qubit can act as a near ideal mirror for (low-intensity) microwave light. Deviation from the ideal result was caused by pure dephasing and qubit decay into nonradiative channels. Although this extra channel is present, the significant extinction of the transmitted signal implies that nonradiative decay is overwhelmed by radiative decay into the line. This is the signature of strong coupling for such a system. Experiments with transmon qubits~\cite{Koch:2007} in the same regime have also been realized by the Chalmers group~\cite{hoi:2011a,Hoi:2012,hoi:2012a}. Interaction of a superconducting qubit with photons propagating in a 1D line has also been studied theoretically~\cite{shen:2005a,shen:2007a,zheng:2010a,peropadre:2012a,rephaeli:2012a,koshino:2012a}.

\begin{figure*}[t] %  figure placement: here, top, bottom, or page
   \includegraphics[width=1.5\columnwidth]{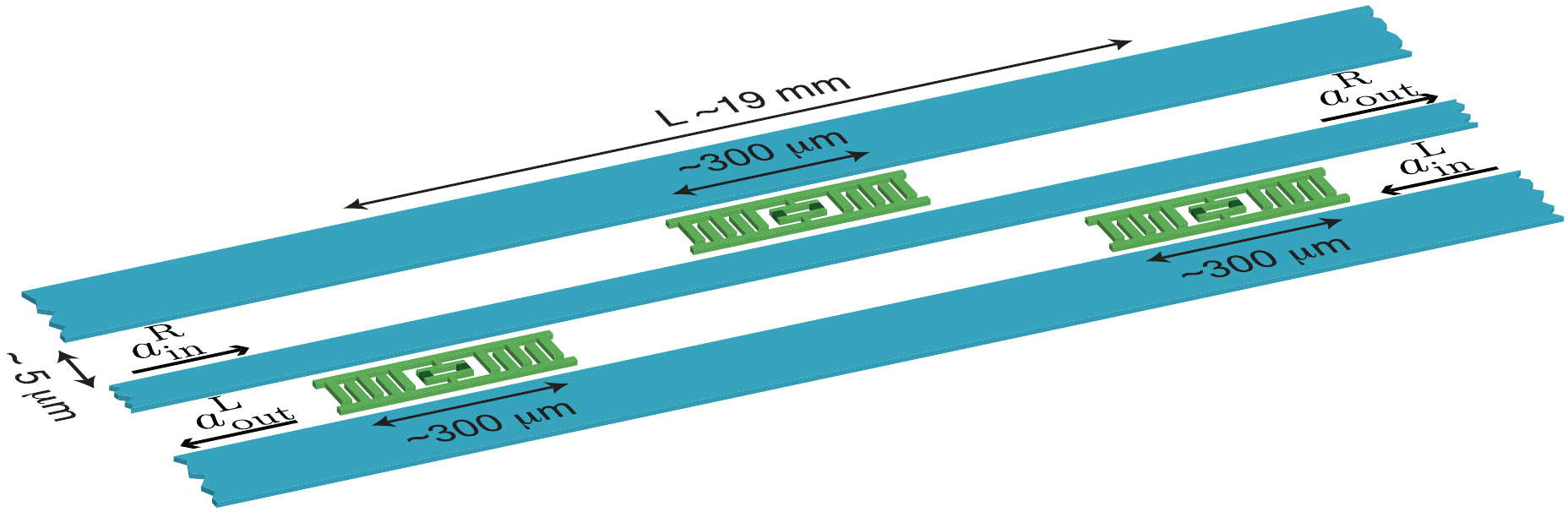} 
   \caption{(Color online) Waveguide QED realization with superconducting circuits: transmon qubits acting as artificial atoms (in green) are coupled to a 1D superconducting transmission line (in blue).}
   \label{fig:systemSchematic}
\end{figure*}

In this paper, we study theoretically the situation where several multi-level superconducting qubits (artificial atoms) interact with the same 1D transmission line. Experimental results on this waveguide QED system are presented in a companion paper~\cite{Loo:2013}.  To understand the main results, it is useful to first consider the well known and simpler case of a single free atom in 3D space. There, interaction of the atom with vacuum fluctuations leads to relaxation, due to emission by the atom of a photon at the atomic transition frequency, and to a Lamb shift of the atomic energy levels, due to emission of virtual photons.  In the presence of a second atom,  real and virtual photons emitted by the first atom can be absorbed by the second, leading to a nontrivial interaction between the two. Therefore, while a single atom acts as a mirror reflecting incident light, two or more atoms will behave in a more complex way~\cite{Lehmberg:1970, Ficek:1990}. However, because of poor mode-matching this interaction is rather weak in 3D.

With near-perfect mode matching, artificial atoms in a 1D transmission line provide an ideal system to study this interaction. Indeed, a single microwave photon could, in principle, travel for several kilometers along a superconducting transmission line before being lost. In this setting the above-mentioned qubit-qubit interaction is therefore long-range and emission of real photons modifies the single-qubit relaxation time~\cite{Ordonez:2004}. As a result, collective decay or, in other words, super- and subradiance will be observed in the presence of several atoms. On the other hand, emission of virtual photons leads to exchange-type interactions between the qubits. Whereas these interactions can be long-range, their characteristics depend on the distance between the qubits. The character of the qubit-qubit interaction therefore changes with the distance between them. To change the distance between the qubits \emph{in situ} is not feasible for a given sample, but changing the wavelength $\lambda$ at which the qubits emit has an equivalent effect. Consequently, it is possible to study the distance dependence of the interaction simply by tuning the qubit transitions frequencies. 

A system of two (artificial) atoms interacting via 3D~\cite{Lehmberg:1970,Agarwal:1974,Lenz:1993,Rudolph:1995,Ficek:2002,Le-Kien:2005,Dzsotjan:2010,Gonzalez-Tudela:2011} and 1D~\cite{Gonzalez-Tudela:2011,Xia:2011,Zueco:2012,Chang:2012a,Shahmoon:2013} open space has been theoretically studied  previously. In particular, non-Markovian effects (which are not relevant to the particular case of interest here) have been studied~\cite{Zheng:2012a}. Here, we adapt Lehmberg's derivation for 3D space~\cite{Lehmberg:1970} to an ensemble of inhomogeneous (artificial) atoms coupled to a 1D transmission line. After eliminating the field degrees of freedom, an effective master equation for the atoms alone is obtained. Some approximations that were reasonable  for atoms in 3D~\cite{Lehmberg:1970} must be revisited. In order to model reflection and transmission of an input beam by the system, we use input-output theory~\cite{collett:1984a}. This allows us to calculate the elastic and inelastic scattering, which we show reveal the effects mentioned above. 

The paper is organized as follows. In Sec.~\ref{sec:WaveguideQED}, we present a theory of waveguide QED. We first present a reduced master equation describing an arbitrary number of many-levels (artificial) atoms coupled to the 1D line and driven by a coherent state. We then apply input-output theory to this system, thereby allowing us to compute elastic and inelastic scattering. The dressed basis, which is useful for understanding collective effects, is discussed. Focusing on the situation where only two qubits are coupled to the line, we show how elastic and inelastic spectra reveal information about collective effects mediated by the open line. This is done in Sec.~\ref{sec:LambdaOver2} considering a $\lambda/2$ separation between the qubits and in Sec.~\ref{sec:LambdaOver4} a $\lambda/4$ separation. We summarize our work in Sec.~\ref{sec:conclusion}. 

In all cases, details of the calculations are relegated to the appendices. In Appendix~\ref{sec:JosephsonJunctions}, we present a derivation of the Hamiltonian for the cases of superconducting transmon qubits coupled to a 1D transmission line. The reduced master equation is derived in Appendix~\ref{sec:masterEquation} and the input-output theory is discussed in Appendix~\ref{sec:inputOutput}. An in-depth discussion of the various approximations, and their validity in the context of waveguide QED with superconducting qubits, is presented in Appendix~\ref{sec:discussionApproximations}. Additional details on the derivation of the reduced master equation can be found in Appendices~\ref{sec:drive},~\ref{sec:coefficientEvaluation}, and~\ref{sec:relaxationDiagonalization}.

\section{Waveguide QED}
\label{sec:WaveguideQED}

\subsection{General effective master equation for the inhomogeneous system}

As illustrated in Fig.~\ref{fig:systemSchematic} with transmon qubits, we consider an ensemble of $N$ inhomogeneous (artificial) atoms, each with $M$ levels. They are dipole coupled to a 1D transmission line. The electromagnetic field in the transmission line can be described by the Hamiltonian~\cite{shen:2005a}
\be
\label{eq:fullHfield}
H_\mathrm{F}=\int_0^\infty d\omega \hbar \omega \left[  \ad_\mathrm{R}(\omega) a_\mathrm{R}(\omega)+\ad_\mathrm{L}(\omega) a_\mathrm{L}(\omega) \right],
\ee
where $\ad_{\mathrm{R}(\mathrm{L})}(\omega)$ creates right- (left-) moving excitations at frequency $\omega$ in the line. The Hamiltonian of the artificial atoms is
\be
\label{eq:fullHatoms}
H_\mathrm{A} = \sum_{j=0}^{N-1} \sum_{m=0}^{M-1} E_{mj} \ket{m_j} \bra{m_j},
\ee 
where $E_{mj}$ is the energy of the $m$th state of the $j$th atom. The interaction Hamiltonian between the line's electric field and the electric dipole for the free artificial atoms can be described as
\be
\label{eq:fullHinteraction}
H_\mathrm{I} = \sum_{j=0}^{N-1}\sum_{m=0}^{M-1}  \hbar g_j \sqrt{m+1} \left(     \Xi_j+\Xi_j^\dagger \right)\sigma_x^{mj}.
\ee
In this expression, $\Xi_j$ is related to the electric field at the location $x_j$ of the $j$th artificial atom,
\be
\label{eq:define_xi}
\Xi_j=-i\int_0^\infty d\omega \sqrt{\omega} \left[ a_\mathrm{L} (\omega) e^{-i \omega x_j/v}+a_\mathrm{R}(\omega) e^{i \omega x_j/v}\right],
\ee
with $v$ the speed of light in the transmission line. We define
\be
\sigma_x^{mj}= \sigma_-^{mj} + \sigma_+^{mj},
\ee
with
\be
\sigma_-^{mj}=\ket{m_j}\bra{(m+1)_j}=\left(\sigma_+^{mj} \right)^\dagger,
\ee
the lowering operator for the $(m+1)$th state of the $j$th atom. The interaction only involves transitions between adjacent states of the atoms, which is a valid approximation for the transmon superconducting qubit behaving as a weakly nonlinear oscillator~\cite{Koch:2007}. Finally, $g_j$ is the (dimensionless) coupling strength between atom $j$ and the field. The expression of $g_j$ for transmon qubits is given in Appendix~\ref{sec:discussionApproximations}.

Following Lehmberg~\cite{Lehmberg:1970}, and as shown in Appendix~\ref{sec:masterEquation}, the effective master equation for the artificial atoms after tracing out the field degrees of freedom can be expressed as~\cite{Lehmberg:1970,Agarwal:1974,Lenz:1993,Ficek:2002,Rudolph:1995,Le-Kien:2005,Dzsotjan:2010,Gonzalez-Tudela:2011,Zueco:2012,Chang:2012a,Zheng:2012a} 
\begin{equation}
\begin{split}
\label{eq:mainMasterEquation}
\dot{\rho}=&-\frac{i}{\hbar}\left[H,\rho\right]\\
&+
\sum_{mj,nk}
\gamma_{mj,nk}\left[\sigma_-^{mj}\rho\sigma_+^{nk}-  \frac{1}{2}\left\{ \sigma_+^{nk} \sigma_-^{mj},\rho \right\}\right],
\end{split}
\end{equation}
with the effective Hamiltonian 
\begin{equation}\label{eq:H}
H = H_\mathrm{A}+\hbar\sum_{mj} d_{mj}(t)\sigma_x^{mj}+\hbar\sum_{mj, nk} J_{mj,nk} \sigma_-^{mj}\sigma_+^{nk}.
\end{equation}
This effective Hamiltonian contains a drive on the atoms proportional to $d_{mj}(t)$. For input coherent states incoming from the left (right) and of frequency $\omega_d$, phase $\theta_{\text{L(R)}}$ and power $P_{\text{L(R)}}$ we show in Appendix \ref{sec:drive} that
\begin{align}\label{eq:drive_amplitude_full_Main}
d_{mj}(t)=&-2\sqrt{\frac{\gamma_{mj,mj}}{2}} \left( \sqrt{\frac{P_\mathrm{L}}{\hbar\omega_{mj}}}\sin\left[ \omega_d (t+t_j+\theta_\mathrm{L})\right] \right.\nnn
&\left.+\sqrt{\frac{P_\mathrm{R}}{\hbar\omega_{mj}}}\sin \left[\omega_d (t-t_j+\theta_\mathrm{R}) \right]\right)
\end{align}
with $t_j=x_j/v$.

Hamiltonian of Eq.~\eqref{eq:H} is Hermitian since $J_{mj,nk}=J_{nk,mj}^*$. As discussed in Appendix~\ref{sec:masterEquation}, in obtaining this expression, we have used the rotating-wave approximation, dropped small non-positive terms in the dissipators and absorbed Lamb shifts in the definition of $H_\mathrm{A}$ (see Appendix~\ref{sec:masterEquation} for the full expression). As seen from Eq.~\eqref{eq:mainMasterEquation}, the effect of the interaction with the transmission line is to  damp atoms at the rate
\begin{align}
\label{eq:relaxationRate}
\gamma_{mj,nk}
=& 2\pi g_k g_j \sqrt{(m+1)(n+1)} 
 \left(\chi_{mjk}+\chi_{nkj}^*\right).
\end{align}
with 
\begin{align}
\chi_{mjk}=&\omega_{mj} e^{i \omega_{mj} t_{kj}},\\
\hbar\omega_{mj}=&E_{m+1,j}-E_{mj}
\end{align}
and $t_{kj}= |x_k - x_j|/v$ the time it takes the  signal to propagate from atom $k$ to atom $j$. For $j = k$, Eq.~\eq{eq:relaxationRate} corresponds to standard relaxation rates of the atoms. As discussed below, for $j \neq k$ this however corresponds to correlated decay. 

The last term of Eq.~\eqref{eq:H} is an exchange interaction between the atoms being mediated by virtual excitations in the line with amplitude
\begin{align}
\label{eq:exchangeInteraction}
J_{mj,nk}
=& -i \pi g_k g_j \sqrt{(m+1)(n+1)}\left( \chi_{mjk}-\chi_{nkj}^*\right).
\end{align}

For the particular case of a pair of levels in two atoms that are tuned to resonance,  $\omega_{mj}=\omega_{nk}$, the expressions for $\gamma_{mj,nk}$ and $J_{mj,nk}$ take a simpler form~\cite{Le-Kien:2005,Dzsotjan:2010,Gonzalez-Tudela:2011,Zueco:2012,Chang:2012a}:
\begin{align}\label{eq:SimplerGamma}
\gamma_{mj,nk}=& 4\pi g_k g_j \omega_{mj} \sqrt{(m+1)(n+1)}
 \cos \left(\omega_{mj}t_{kj}\right),
\end{align}
and
\begin{align}\label{eq:SimplerJ}
J_{mj,nk}=&2 \pi g_k g_j \omega_{mj}  \sqrt{(m+1)(n+1)}\sin \left(\omega_{mj}t_{kj}\right).
\end{align}
This form makes it clear that the magnitude of these two quantities has an oscillatory dependence on interatomic separation.

\subsection{Input-output theory}

To compare theoretical predictions to experiments measuring reflection and transmission of light by the system, we derive in Appendix~\ref{sec:inputOutput} the input-output boundary condition in the presence of artificial atoms coupled to the line. Only the main results are presented in this section. Following the standard prescription~\cite{collett:1984a}, we find
\begin{equation}\label{eq:aoutR}
a_{\text{out}}^\mathrm{R}(t) = a_{\text{in}}^\mathrm{R}(t)+\sum_{mj}  e^{-i\omega_{mj} t_j} \sqrt{\frac{\gamma_{mj,mj}}{2}}\sigma_-^{mj}
\end{equation}
and
\begin{equation}\label{eq:aoutL}
a_{\text{out}}^\mathrm{L}(t)  = a_{\text{in}}^\mathrm{L}(t)+\sum_{mj}  e^{i\omega_{mj} t_j} \sqrt{\frac{\gamma_{mj,mj}}{2}}\sigma_-^{mj},
\end{equation}
where 
\be
a_{\text{in}}^\mathrm{R}(t)=\int_0^\infty \frac{d\omega}{\sqrt{2\pi}} a_\mathrm{R}(\omega,t_0) e^{-i \omega t}
\ee 
represents the input field arriving at the atoms from the left and
\be
a_{\text{out}}^\mathrm{R}(t)=\int_0^\infty \frac{d\omega}{\sqrt{2\pi}} a_\mathrm{R}(\omega,t_1) e^{-i \omega t}
\ee
the output field propagating to the right after interaction with the system. As is standard in input-output formalism, $t_0<t$ and $t_1>t$ refer to a time respectively before or after interaction with the system. Similar expressions can be found for the left-moving fields. 

Assuming for example that the system is driven from the left, it is possible to compute using Eqs.~\eqref{eq:aoutR} and \eqref{eq:aoutL} the transmission coefficient
\begin{equation}\label{eq:T}
|t|^2=|\mean{\aout^\mathrm{L}}/\mean{\ain^\mathrm{L}}|^2
\end{equation} 
and the reflection coefficient
\begin{equation}\label{eq:R}
|r|^2=|\mean{\aout^\mathrm{R}}/\mean{\ain^\mathrm{L}}|^2
\end{equation}
corresponding to elastic scattering. Another useful quantity is the power spectrum of the output field
\be\label{eq:DefSomega}
S^{\alpha}[\omega] = \int_{-\infty}^\infty dt e^{i\omega t} \langle a_{\text{out}}^{\alpha\dag} (t) a_{\text{out}}^{\alpha}(0) \rangle
\ee
for $\alpha = \mathrm{R}, \mathrm{L}$ and which corresponds to inelastic scattering for $\omega\neq0$.  Analytical or numerical predictions for both the elastic and inelastic scattering will be presented below for two choices of interatomic separations.

\subsection{Dressed basis}
\label{sec:DressedBasis}

Elsewhere~\cite{Loo:2013}, we report measurements of $|t|^2$, $|r|^2$, and $S^{\alpha}[\omega]$ for $N=2$ transmons coupled to the line. There, quantitative agreement with numerical calculations is presented. Here, we take $M=N=2$ in the reduced master equation~\eqref{eq:mainMasterEquation} and focus on the analytical results. Quantitative agreement between the theoretical description below and the experimental results of Ref.~\cite{Loo:2013} can be obtained.  

In a frame rotating at drive frequency $\omega_d$, taking $N=M=2$ leads to
\begin{align}
\label{eq:reduced_me_22}
\dot{\rho}=&-\frac{i}{\hbar}\left[H,\rho\right]+\sum_{jk}\gamma_{jk}\left[\sigma_-^{j}\rho \sigma_+^{k}-  \frac{1}{2}\left\{ \sigma_+^{k} \sigma_-^{j},\rho \right\}\right],
\end{align}
where
\begin{equation}\label{eq:Heff:22}
\begin{split}
H/\hbar=
& \sum_{j}  \Delta_{j}\ket{e_j}\bra{e_j}+\sum_{j} \left(  \epsilon_{j} \sigma_+^{j}+\hc\right)\\
&+ J (\sigma_-^{1}\sigma_+^{2} + \sigma_+^{1}\sigma_-^{2}),
\end{split}
\end{equation}
with $\ket{e_j}$ the excited state of qubit $j$, $\Delta_{j} = \omega_{0j} - \omega_d$, $J=J_{0j,0k}$ and $\gamma_{jk}=\gamma_{0j,0k}+\gnrj\delta_{jk}$. The rate  $\gnrj$ represents nonradiative decay of qubit $j$. In practice, it is easy to realize a situation where qubit decay will be dominated by emission into the line, that is $\gamma_{jj}\gg \gnrj$. In this open-line setting, satisfying this inequality corresponds to the strong-coupling regime~\cite{Abdumalikov:2011}. Assuming that the qubits are driven from the left only,  Eq.~\eq{eq:drive_amplitude_full_Main} for the  drive amplitude now takes the simpler form
\begin{align}
\epsilon_{j}
=
-i \sqrt{\frac{\gamma_{0j,0j}\omega_d}{2\omega_{0j}}}  \mean{\ain^\mathrm{L}}
e^{-i \omega_d t_j}.
\end{align}

To deal with correlated decay described by the last term of Eq.~\eq{eq:reduced_me_22}, it is useful to move to a basis that diagonalizes the dissipation matrix with components $\gamma_{j,k}$. As shown in  Appendix \ref{sec:relaxationDiagonalization}, this leads to the more standard form for the last term of Eq.~\eqref{eq:reduced_me_22}, which now reads
\begin{align}
\sum_{\mu =B,D} \Gamma_\mu \mathcal D \left[ \sigma_-^\mu \right]\rho,
\end{align}
where $\mathcal D[x] \rho=x \rho x^\dagger-\left\{x^\dagger x,\rho \right\}/2$ is the standard dissipator that is now acting on the dressed lowering operators
\be
\label{eq:sigmaHL}
\sigma_-^{\mu}=\frac{\left(\Gamma_{\mu}-\gamma_{11} \right) \sigma_-^0 + \gamma_{01}^*\sigma_-^1}{\sqrt{\left(\Gamma_{\mu}-\gamma_{11} \right)^2+|\gamma_{01}|^2}},
\ee
with $\mu=B,D$ and with correlated decay rates
\be
\Gamma_{B/D}=\frac{\gamma_{00}+\gamma_{11}}{2}\pm \sqrt{\left(\frac{\gamma_{00}-\gamma_{11}}{2}\right)^2+|\gamma_{01}|^2}.
\ee
The subscripts $B$ and $D$ refer to bright and dark respectively. Due to dependence on the qubit separation, both the correlated qubit decay $\Gamma_{\mu}$ and exchange interaction $J$ can be tuned by a modification of the qubit's transition frequency. 

Below we consider the case of two qubits tuned in resonance at a frequency $\omega_0$ such that the distance $d$ between them corresponds to $\lambda_0$ or $3\lambda_0/4$, with $\lambda_0 = 2\pi v/\omega_0$. In both cases, reflection and transmission coefficients are calculated as well as the corresponding power spectra.

\section{$\bm{\lambda/2}$ separation: Sub- and Superradiance}
\label{sec:LambdaOver2}

\subsection{Discussion}

We first consider a pair of qubits tuned in resonance at a frequency $\omega_{0}$, whose associated wavelength $\lambda_{0}$ is equal to $d = \lambda_{0} / 2$. To simplify the discussion, we let $\gnr \equiv \gnr^0 \sim \gnr^1$. In the strong-coupling regime, $\gnrj  \ll \gamma_{00},\gamma_{11}$, and the nonradiative relaxation rate is a small perturbation. The above assumption that the nonradiative rates are equivalent for both qubits will therefore not affect the results much. 

With this simplification and the choice $d = \lambda_{0} / 2$, the off-diagonal decay rate $\gamma_{01}$ defined in Eq.~\eq{eq:relaxationRate} can be written as
\begin{align}
\gamma_{01} = \pm \sqrt{(\gamma_{00}-\gnr) (\gamma_{11}-\gnr)}.
\end{align}
This leads to
\be\label{eq:GammaDarkBrightLambdaOver2}
\Gamma_{D}= \gnr
\ll
\Gamma_{B}=\gamma_{00}+\gamma_{11}-\gnr.
\ee
In other words, for $d = \lambda_{0} / 2$ the state $\ket{D}$ defined by $\sigma_-^{D} \ket{D}  = 0$ is dark as its decay rate is purely nonradiative. On the other hand, the state $\ket{B}$ defined by $\sigma_-^{B} \ket{B} = 0$ is bright. This corresponds, respectively, to sub- and superradiance~\cite{Lenz:1993,Le-Kien:2005,Dzsotjan:2010,Gonzalez-Tudela:2012}. Moreover, for this half-wavelength setting, the exchange interaction is absent with $J=0$.
\begin{figure}[t] %  figure placement: here, top, bottom, or page
   \includegraphics[width=\columnwidth]{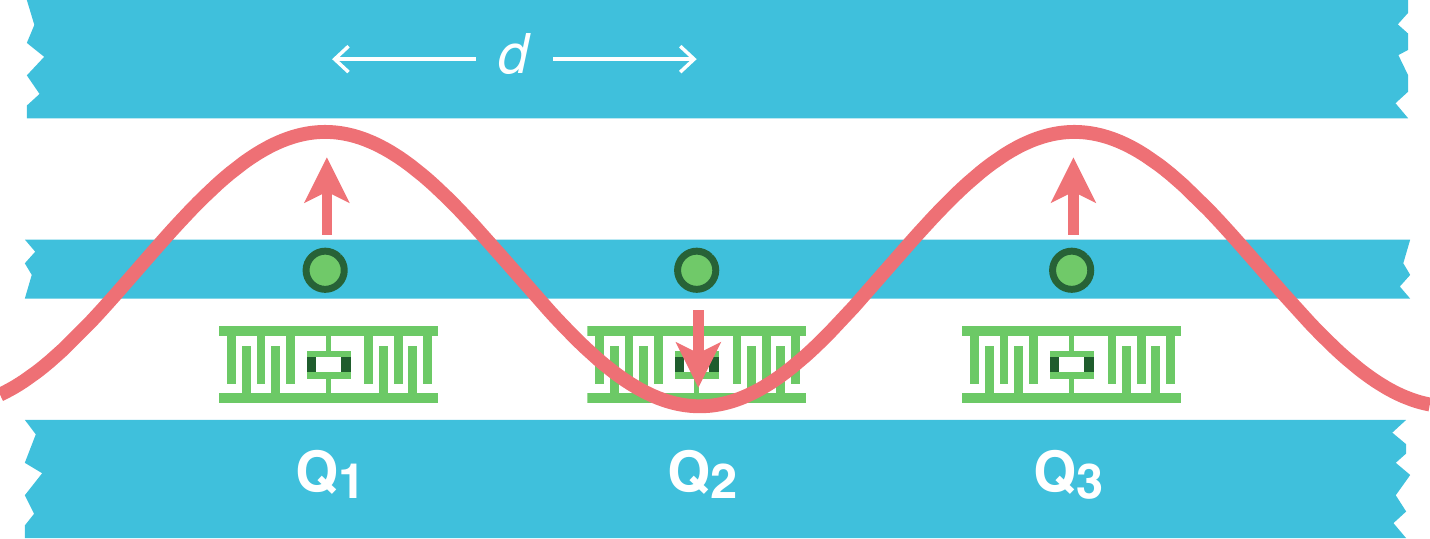} 
   \caption{(Color online) Schematic representation of three transmon qubits in a 1D transmission line. Qubits are considered as point-like objects and their locations $x_j$ along the line are represented by circles. $Q_1$ and $Q_2$ are separated by $\lambda_0/2$ while $Q_1$ and $Q_3$ by $\lambda_0$. If $Q_1$ and $Q_3$ are identical, only symmetric superpositions of these two qubits can be excited by an external drive of wavelength $\lambda_0$. On the other hand, since they are separated by $\lambda_0/2$, only antisymmetric superpositions of $Q_1$ and $Q_2$ can be excited.}
   \label{fig:LambdaOver2}
\end{figure}

That $\ket{B}$ and $\ket{D}$ are bright and dark, respectively, can also be seen from the Hamiltonian. Indeed, by inverting Eq.~\eqref{eq:sigmaHL}, it is possible to rewrite the driving term in Eq.~\eqref{eq:Heff:22} as
\be
\sum_{\mu = B,D}\hbar\left(\epsilon_\mu \sigma_+^\mu+\hc\right).
\ee
For $|\Delta_j|/\omega_0 \ll 1$, which is easily satisfied,  the drive amplitudes now take the forms
\begin{equation}\label{eq:epsilon:LambdaOver2}
\begin{split}
\epsilon_{D} & \approx 0,
\\
\epsilon_{B} & \approx -i  \langle a_{\text{in}}^\mathrm{L} \rangle e^{-i \omega_d t_0} \sqrt{\frac{\gamma_{00}+ \gamma_{11}}{2} -\gnr}.
\end{split}
\end{equation}
Clearly, $\ket{D}$ cannot be driven from the ground state $\ket{gg}$. This can be understood intuitively from Fig.~\ref{fig:LambdaOver2} in the case $\gamma_{00}=\gamma_{11}$. First, only consider the two leftmost qubits, $Q_1$ and $Q_2$. As illustrated in Fig.~\ref{fig:LambdaOver2}, when driven on resonance, $Q_1$ and $Q_2$ experience opposite phases of the driving field as they are separated by $d = \lambda_{0} / 2$. In this case, transitions between $\ket{gg}$ and $\ket{B}=(\ket{ge} - \ket{eg})/\sqrt{2}$ are allowed while transitions between $|gg\rangle$ and $\ket{D}= (\ket{ge} + \ket{eg})/\sqrt{2}$ are forbidden.  These selection rules are captured by Eq.~\eqref{eq:epsilon:LambdaOver2} and are akin to what is observed in circuit QED in the presence of  two qubits in the same resonator~\cite{majer:2007a,filipp:2011a}.  As damping, just like driving, is an interaction of the qubits with the line, we also find in Eq.~\eqref{eq:GammaDarkBrightLambdaOver2} that $\Gamma_D = \gnr$ or, in other words, that $\ket D$ does not decay radiatively. As expected from these simple arguments, the situation is reversed for $Q_1$ and $Q_3$ in Fig.~\ref{fig:LambdaOver2}, which are separated by $d = \lambda_{0}$. In this case, $\ket{B}=(\ket{ge} + \ket{eg})/\sqrt{2}$ and $\ket{D}= (\ket{ge} - \ket{eg})/\sqrt{2}$.

\begin{figure}[t] %  figure placement: here, top, bottom, or page
   \includegraphics[width=\columnwidth]{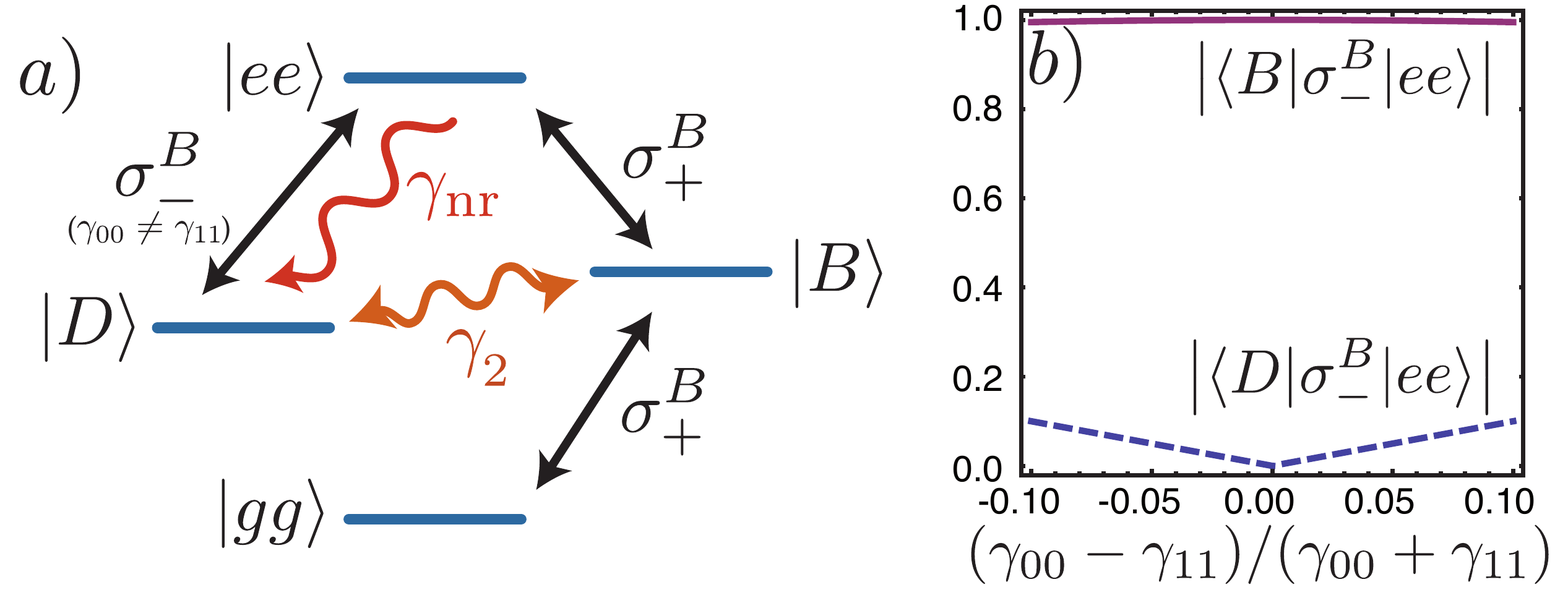} 
   \caption{(Color online) (a) Schematic energy level diagram for two qubits in a transmission line. The dark state $\ket{D}$  cannot be driven directly from the ground state $\ket{gg}$. If the qubits are identical, it can however be excited by nonradiative relaxation $\gnr$ from $\ket{ee}$. In the situation where the qubit relaxation rates are different ($\gamma_{00}\neq\gamma_{11}$), it can be excited indirectly from $\ket{B}$ via $\ket{ee}$. Dephasing ($\gamma_2$) can also cause transition between $\ket{B}$ and $\ket{D}$. (b) Matrix elements of $\sigma_-^B$ as a function of the relaxation rate asymmetry  $(\gamma_{00}-\gamma_{11})/(\gamma_{00}+\gamma_{11})$.
}
   \label{fig:LambdaOver2Levels}
\end{figure}

It is important to point out that, since $\left[ \sigma^{B/D}_\pm,\sigma^{B/D}_\mp\right]\neq 0$ and $\left[ \sigma^{B/D}_\pm,\sigma^{D/B}_\mp\right] \neq 0$, $\ket{D}$ will not be completely dark in practice and especially not in the presence of finite nonradiative decay $\gnr$. Indeed, as illustrated in Fig.~\ref{fig:LambdaOver2Levels}(a), the joint excited state $\ket{ee}$ can be reached from $\ket{B}$ by driving with $\sigma_+^B$. From this state, $\ket{D}$ can be populated with the action of $\sigma_-^B$ when $\gamma_{00} \neq \gamma_{11}$. This is because the matrix element of $\sigma_-^B$ between $\ket{ee}$ and $\ket{D}$ is proportional to the asymmetry $(\gamma_{00} - \gamma_{11})/(\gamma_{00} + \gamma_{11})$, as shown in Fig.~\ref{fig:LambdaOver2Levels}(b).  The dark state can also be populated by nonradiative decay. It is interesting to point out that, while nonradiative relaxation cannot be controlled in this system, the indirect driving of $\ket{D}$ from $\ket{ee}$ can be tuned by controlling the asymmetry between $\gamma_{00}$ and $\gamma_{11}$, something that can be done by tuning the qubit frequency.

\subsection{Elastic scattering}
\label{sec:symmetricSubradiance}

In this section, we compute the elastic scattering or, more precisely, the transmission, Eq.~\eqref{eq:T}, and reflection coefficients, Eq.~\eqref{eq:R},  assuming the two-qubit system to be driven from the left. To simplify the discussion, as before we take both qubits to have the same frequency, such that $\Delta=\Delta_j$, and have the same total decay rate $\gamma=\gamma_{jj}$. It is also useful to introduce $\gr=|\gamma_{01}|=\gamma-\gnr$, the radiative contribution to the decay rate. With these definitions and Eq.~\eqref{eq:aoutR}, we find for the outgoing field that
\begin{align}
a_{\text{out}}^{\mathrm{R/L}}&=a_{\text{in}}^{\mathrm{R/L}}\pm i  \sqrt{\gr} \sigma_-^B.
\end{align}
Solving for $\langle \sigma_-^B \rangle$ in steady state using the master equation \eqref{eq:reduced_me_22}, we have, to first order in the drive amplitude $\mean{\ain^\mathrm{L}}$,
\begin{align}
\label{eq:aoutRLambdaOver2}
\mean{\aout^\mathrm{R}} &=\mean{\ain^\mathrm{L}}\frac{-i (\Gamma_B-\gnr)/2}{\Delta -i\Gamma_B/2},\\
\label{eq:aoutLLambdaOver2}
\mean{\aout^\mathrm{L}} &=\mean{\ain^\mathrm{L}}\left(1-\frac{-i(\Gamma_B-\gnr)/2}{\Delta-i\Gamma_B/2}\right).
\end{align}
These equations are expected for a single qubit with relaxation rate $\Gamma_B$~\cite{Astafiev:2010}. Indeed, in the absence of nonradiative decay, there is full extinction of the transmission and complete reflection when driving (with low power) on resonance $\Delta = 0$. While we are dealing here with a four-level system, this simple behavior is observed because, as illustrated in Fig.~\ref{fig:LambdaOver2Levels}, at low irradiation power and in the absence of nonradiative decay or additional dephasing, only two levels are relevant, $\{\ket{gg},\ket{B}\}$. The two qubits behave like a single two-level system coupled to the line and with decay rate $\Gamma_B$.

\begin{figure}[tb] %  figure placement: top, bottom
   \includegraphics[width=0.85\columnwidth]{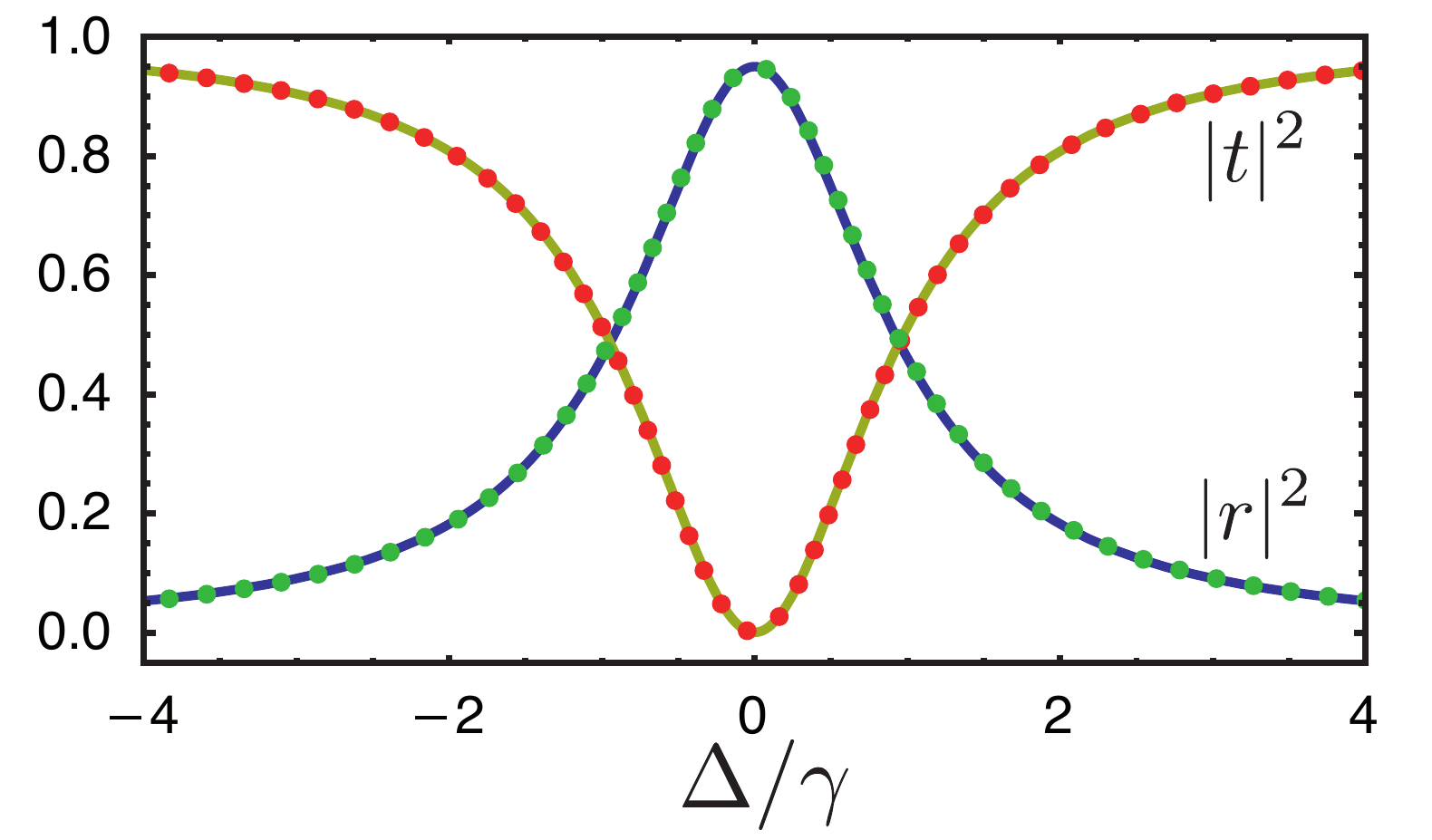} 
   \caption{(Color online) Transmission and reflection coefficients as a function of normalized detuning $\Delta/\gamma$ between the qubits equal transition frequencies and drive. The qubits are tuned such that they are separated by $d=\lambda_0$. There is full transmission extinction on resonance. The width of both $|t|^2$ and $|r|^2$ is given by the superradiant rate $\Gamma_B \sim 2 \gamma$. Solid lines are obtained from analytical results whereas the dotted lines are obtained from numerical simulations of the reduced master equation. The selected parameters are $\gr=0.95 \gamma$ and $\gamma/2\pi = 18.8$ MHz.} 
   \label{fig:aoutLambdaOverTwo}
\end{figure}

The output fields can also be obtained exactly analytically, but this leads to expressions that are too long to be worth reproducing here. The transmission $|t|^2$ and reflection coefficients $|r|^2$ obtained from these exact expressions are illustrated in Fig.~\ref{fig:aoutLambdaOverTwo}. These are in excellent agreement with results obtained from numerical integration of the reduced master equation Eq.~\eqref{eq:mainMasterEquation}. As expected from the above discussion, the width of the transmission dip is given by the superradiant rate $\Gamma_B \sim 2 \gamma$. %

\subsection{Inelastic scattering}

\begin{figure}[t] %  figure placement: here, top, bottom, or page
   \includegraphics[width=\columnwidth]{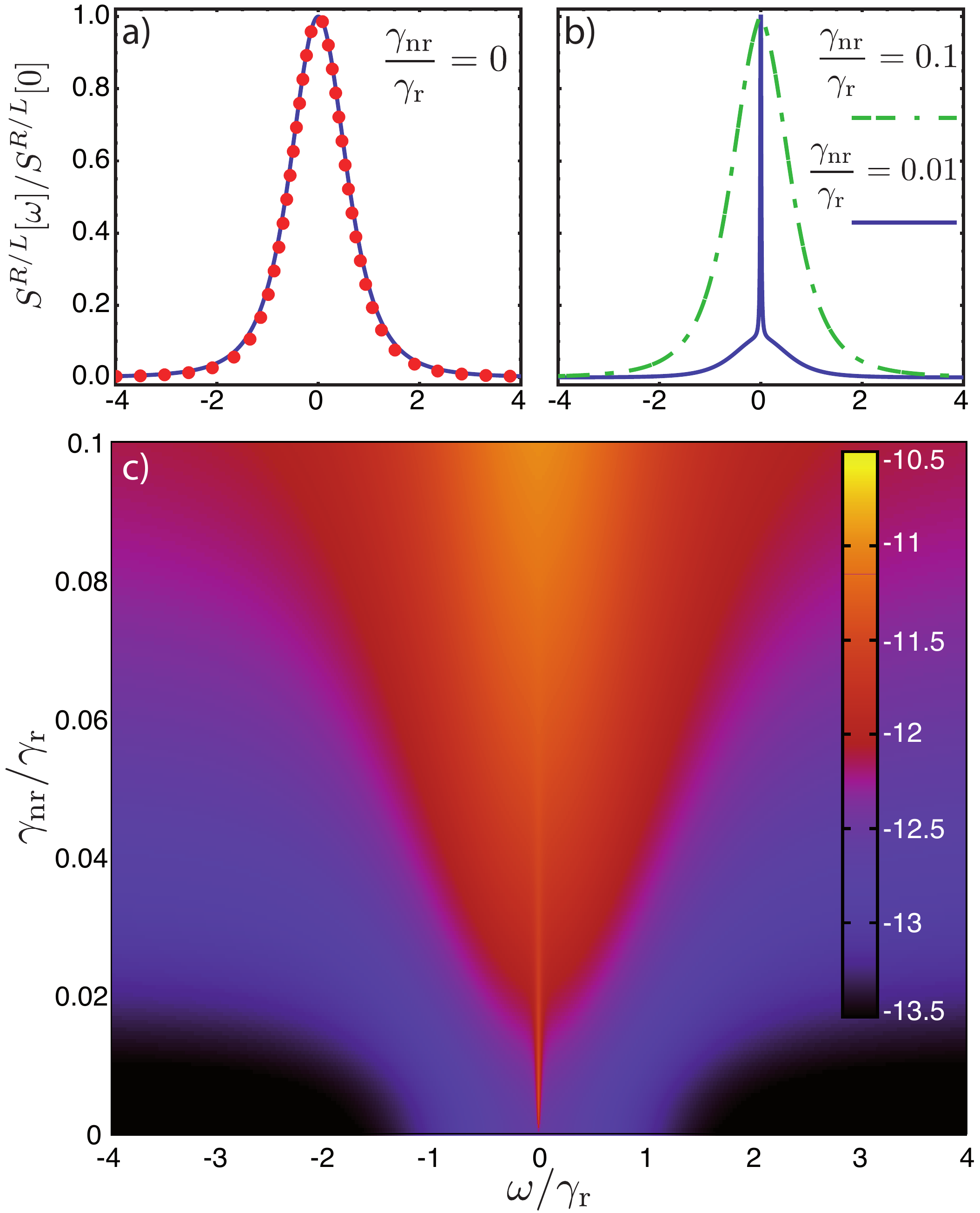} 
   \caption{(Color online) Power spectral density $S^\alpha[\omega]$ as a function of normalized frequency $\omega/\gr$ and for a weak coherent tone corresponding to an amplitude $\epsilon_B/\gr=0.005 $. The qubits are tuned such that they are separated by $d=\lambda_0$. In the absence of nonradiative relaxation or asymmetry in the qubit decay rates, a squared Lorentzian of width $\Gamma_B=2 \gr$ is observed. In the presence of nonradiative relaxation the dark state can be populated and a narrow peak appears in the spectrum.  (a)~Analytical (solid blue line) and numerical (red dots) power spectral densities for $\gnr =0$. (b)~Numerical power spectral density for $\gnr/\gr =0.1$ (green dashed line) and $\gnr/\gr=0.01$ (solid blue line). (c)~Log$_{10}$ of the numerical power spectral density vs frequency and as a function of nonradiative relaxation $\gnr/\gr$. $\gr/2\pi = 17.9$ MHz.}
   \label{fig:spectrumLambdaOverTwoGammaNR}
\end{figure}

As argued above, for $\gnr/\gr=0$ and in the absence of pure dephasing, the state $\ket{D}$ is unpopulated and at low enough power, we are left with an effective two-level system $\{\ket{gg},\ket{B}\}$. In this case, the normalized power spectral density takes the simple form~\cite{walls:2008a}
\be
S^{\mathrm{R}/\mathrm{L}}[\omega] 
= \frac{8{\epsilon_B^\text{eff}}^4}{[(\Gamma_B/2)^2+\omega^2]^2},
\ee
with $\epsilon_B^\text{eff}$ the effective driving strength of the superradiant state. Figure~\ref{fig:spectrumLambdaOverTwoGammaNR}(a) plots this expression with $\epsilon_B^\text{eff}$ evaluated as a fit parameter for results obtained from numerical integration of the master equation \eqref{eq:mainMasterEquation} and Eq.~\eqref{eq:DefSomega} in the limit $\gnr/\gr=0$. As illustrated in Fig.~\ref{fig:spectrumLambdaOverTwoGammaNR}(b), for $\gnr/\gr$ finite but small, results obtained from numerical integration deviate from the above simple expression and show a sharp peak in the spectral density centered at zero (solid blue line). This is a signature of the subradiant state $\ket{D}$ that can become populated, as illustrated schematically in Fig.~\ref{fig:LambdaOver2Levels}, via $\ket{ee}$ by nonradiative decay $\gnr$ and via $\ket{B}$ by dephasing $\gamma_2=\gnr/2 + \gamma_\varphi$. This peak should not be confused with the Rayleigh-scattered radiation which results in a $\delta$ peak at $\omega = 0$ and which we have removed here. 

As illustrated in Fig.~\ref{fig:spectrumLambdaOverTwoGammaNR}(b), the signature of the dark state disappears in the presence of large nonradiative decay. The evolution of this feature as a function of $\gnr/\gr$ is presented in Fig.~\ref{fig:spectrumLambdaOverTwoGammaNR}(c). Whereas a finite nonradiative decay rate is useful to observe both the signature of super- and subradiance, it is required for the system to be in the strong-coupling regime for both features to be observable. As shown in Ref.~\cite{Loo:2013}, this can be achieved with transmon qubits. 

\begin{figure}[t] %  figure placement: here, top, bottom, or page
   \includegraphics[width=\columnwidth]{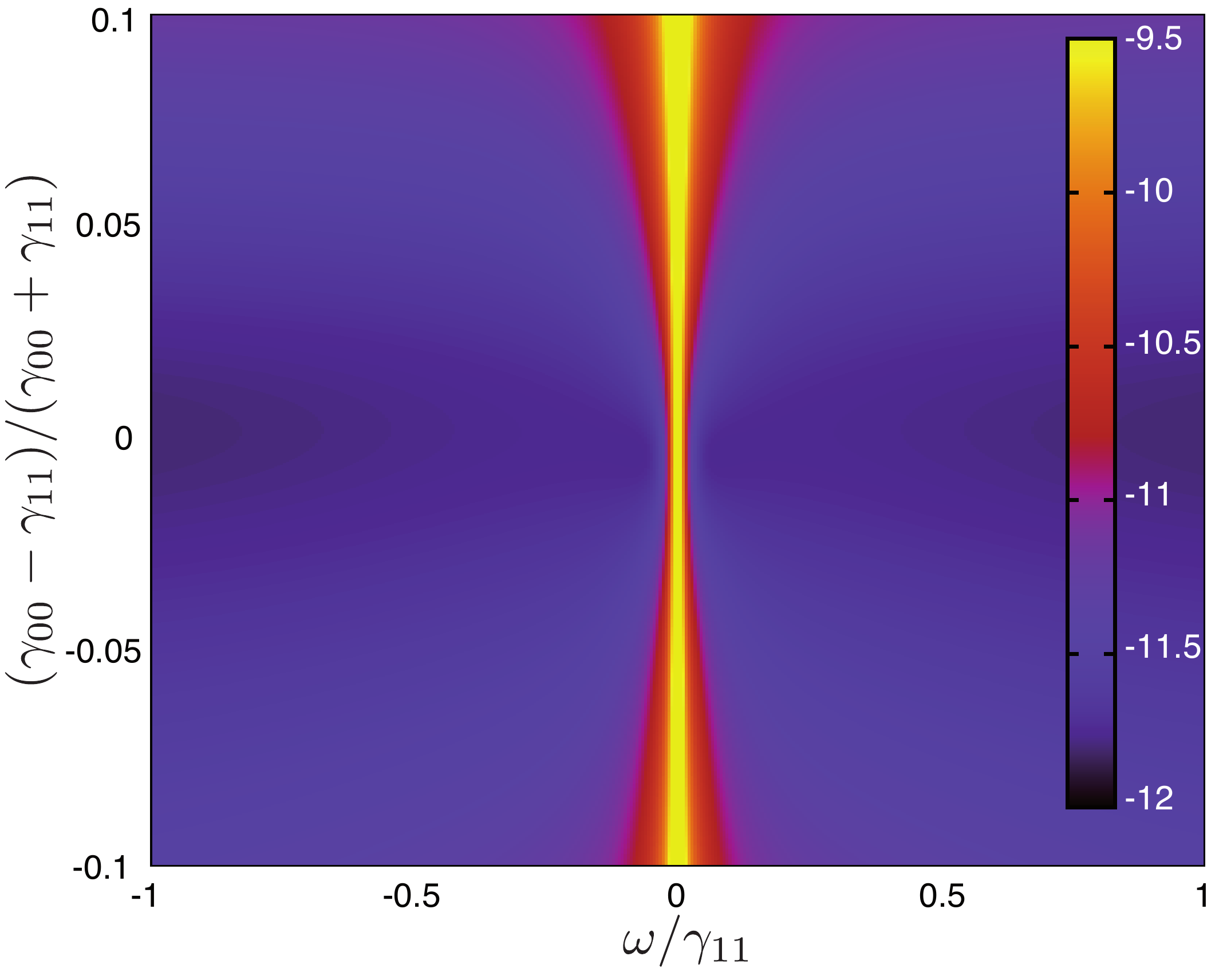} 
   \caption{(Color online) Log$_{10}$ of the numerical power spectral density $S^\alpha[\omega]$ as a function of normalized frequency $\omega/\gamma_{11}$ and relaxation rate asymmetry  $(\gamma_{00}-\gamma_{11})/(\gamma_{00}+\gamma_{11})$. The drive power and nonradiative decay are chosen such that $\epsilon_B/\gamma_{11}=0.005 $ and $\gnr/\gamma_{11}=0.01$, corresponding to a weak coherent drive tone in the strong-coupling limit. The qubits are tuned such that they are separated by $d=\lambda_0$.  Asymmetry between the relaxation rates opens a new drive channel for the dark state $\ket{D}$, causing power broadening. $\gamma_{11}/2\pi = 18.1$ MHz.}
   \label{fig:spectrumLambdaOverTwoAsymmetry}
\end{figure}

The above results have been obtained in the idealized case where $\gamma_{00}=\gamma_{11}$. These decay rates, defined below Eq.~\eqref{eq:Heff:22}, contain both the radiative and the nonradiative contributions. Some asymmetry in the decay rates is to be expected in practice. As illustrated in Fig.~\ref{fig:LambdaOver2Levels}, this leads to a finite transition matrix element between $\ket{ee}$ and the dark state $\ket{D}$. The effect of this asymmetry is illustrated in Fig.~\ref{fig:spectrumLambdaOverTwoAsymmetry}, which presents the numerically computed power spectral density as a function of both frequency and asymmetry $(\gamma_{00}-\gamma_{11})/(\gamma_{00}+\gamma_{11})$. These results are obtained for a constant $\gnr/\gamma_{11}=0.01$ corresponding to the strong coupling limit. This additional population mechanism for the dark state leads to power broadening of the sharp feature centered around $\omega = 0$. However, with up to 10\% asymmetry, this signature of superradiance is expected to be clearly observable at low power. This is confirmed experimentally~\cite{Loo:2013}.

\section{$\bm{\lambda/4}$ separation: Exchange interaction}
\label{sec:LambdaOver4}

\subsection{Discussion}

\begin{figure}[tb] 
   \includegraphics[width=\columnwidth]{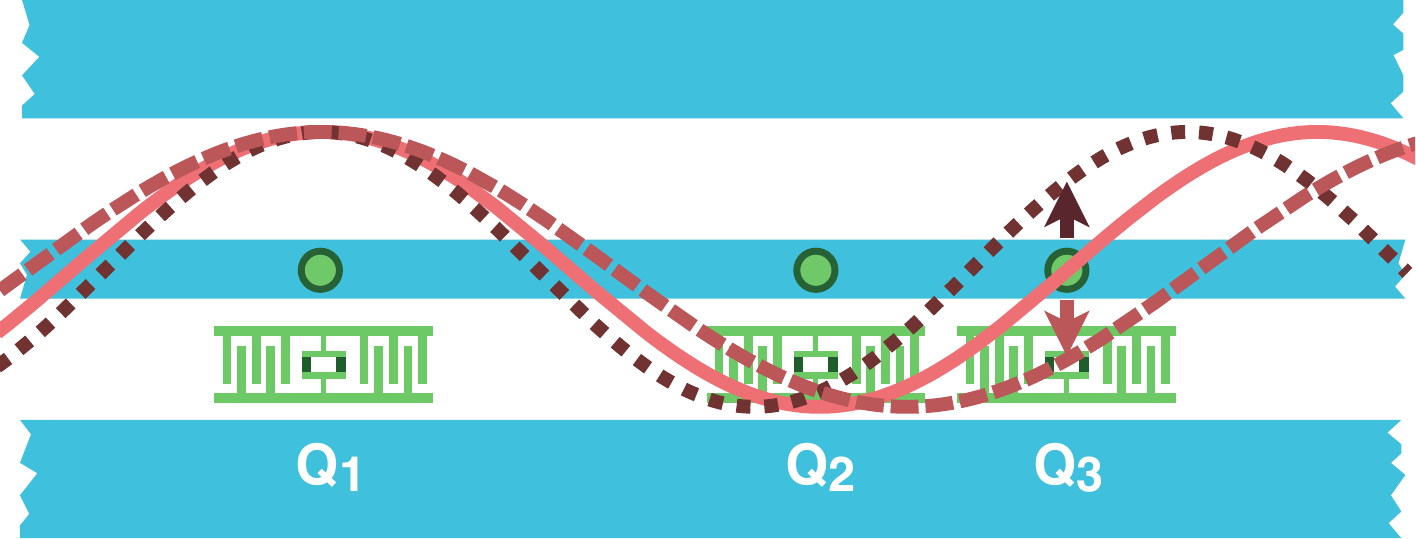} 
   \caption{(Color online) Schematic representation of three transmon qubits in a 1D transmission line. Qubits are considered as point-like objects and their location $x_j$ along the line is represented by circles. As illustrated by the solid line, the distance between $Q_1$ and $Q_3$ corresponds to $3\lambda_0/4$. At the location of $Q_3$, modes of frequency around $3\lambda_0/4$ have opposite signs (see dashed and dotted line). On the other hand, for a separation corresponding to $\lambda_0/2$ just like $Q_1$ and $Q_2$, all modes have the same sign around $Q_2$.}
   \label{fig:LambdaOver4}
\end{figure}

We now consider the situation where the transition frequency of both qubits is chosen such that the qubit separation $d$ is an odd multiple of $\lambda_{0}/4$. This is illustrated in Fig.~\ref{fig:LambdaOver4} where $Q_1$ and $Q_3$ are separated by $3\lambda/4$. As can be seen from Eqs.~\eqref{eq:SimplerGamma} and \eqref{eq:SimplerJ}, in this case the correlated decay rate $\gamma_{01}\propto \cos(2\pi d/\lambda_0)$ is zero and the exchange interaction $\propto \sin(2\pi d/\lambda_0)$ takes its maximal value $|J|= \gr/2$. 

That this interaction is at a maximum for this separation can be understood intuitively from Fig.~\ref{fig:LambdaOver4} and by going back to the origin of the virtual interaction term in the derivation of the effective master equation. Indeed, as can be seen in detailed calculation presented in Appendix~\ref{sec:masterEquation}, the exchange interaction $J$ is a modification of the Lamb shift in the presence of multiple qubits coupled to the line. Basically, virtual photons  emitted and reabsorbed by a given qubit contribute to the qubit's Lamb shift. In the presence of two (or more) qubits, virtual photons can be emitted by one qubit and absorbed by the other, leading to an effective qubit-qubit interaction. This type of exchange interaction is well known in circuit QED where the qubits interact strongly with a single mode of a resonator leading to $J=g_1 g_2/\delta$, with $\delta$ the detuning of both qubits to the resonator~\cite{majer:2007a,filipp:2011a}. In the present open-line context where the qubits interact with a continuum of modes, $J$ is of the same form but is now an integral over all continuous modes \emph{except} the continuous modes lying at qubit transition frequency~\footnote{The exclusion of the qubit transition frequency comes from the principal part in Eq.~\protect\eqref{eq:usefulIdentity}. Indeed, in this equation, the Dirac delta is responsible for relaxation (by emission of photons at the qubit transition frequency) while the principal part is responsible for the Lamb shift and the exchange interaction.}. 

As illustrated in Fig.~\ref{fig:LambdaOver4} for $Q_1$ and $Q_3$, the continuous modes at longer wavelength than $3\lambda_0/4$ (dashed line) have a phase of opposite sign at the location of the second qubit with respect to continuous-modes of shorter wavelength than $3\lambda_0/4$ (dotted line). Moreover, since these continuous modes are, respectively, below and above the qubit frequency, their respective detuning $\delta$ is also of opposite sign. This double change of sign results in a finite exchange interaction because the contribution to $J$ of the modes around $3\lambda_0/4$ all have the same overall (negative) sign. In contrast, for $Q_1$ and $Q_2$ which are separated by $\lambda_0/2$, the phases of all the continuous-modes at $Q_2$ have the same sign while the detuning $\delta$ changes sign. In this case, the exchange interaction vanishes when integrating over all continuous-modes above and below $\lambda_0/2$.

Assuming that the qubits are in resonance, and taking $\gamma=\gamma_{jj}$ for simplicity, this discussion can be made more formal by working in the dressed basis, which diagonalizes the effective Hamiltonian, Eq.~\eqref{eq:Heff:22}. In this situation, the dressed lowering operators,  Eq.~\eqref{eq:sigmaHL}, take the simple form 
\be\label{eq:SigmaMBD}
\sigma^{B/D}_- =  \frac{\sigma^1_- \pm \sigma^0_-}{\sqrt{2}}.
\ee
The master equation then reads
\be
\dot{\rho}=-\frac{i}{\hbar}\left[ H,\rho \right] + \gamma \sum_{i=B,D} \mathcal D\left[ \sigma_-^i \right] \rho,
\ee
where 
\be\label{eq:HeffLambdaOver4}
H=\sum_{i=B,D} \hbar \omega_i \sigma_+^i \sigma_-^i+\sum_{i=B,D}\hbar \left(\epsilon_i \sigma_+^i+\hc \right),
\ee
and $\omega_{B/D}=\Delta \pm J$, $\epsilon_{B/D}=(\epsilon_1 \pm\epsilon_0)/\sqrt{2}$. As expected, in the dressed basis, the system is described by two driven eigenstates whose frequencies differ by $2J$.

\subsection{Elastic scattering}

We now turn to elastic scattering. Using Eqs.~\eqref{eq:aoutL} and \eqref{eq:aoutR}, the output fields can be expressed as
\begin{align}
\label{eq:aoutLambdaOverFour1}
a_{\text{out}}^\mathrm{R}(t)&=\sqrt{\frac{\gr}{2}}e^{-i\omega_{0} t_0} \left[   \sigma_-^{0} -i \sign{J} \sigma_-^{1}\right],\\
\label{eq:aoutLambdaOverFour2}
a_{\text{out}}^\mathrm{L}(t)&=a_{\text{in}}^\mathrm{L}+\sqrt{\frac{\gr}{2}} e^{i\omega_{0} t_0} \left[   \sigma_-^{0}+i \sign{J} \sigma_-^{1}\right].
\end{align}
To first order in the drive amplitude $\mean{\ain^\mathrm{L}}$, we then find the expectation values of these two quantities:
\begin{align}
\label{eq:aoutLLambdaOver4}
\mean{\aout^\mathrm{L}}
&= 
\mean{\ain^\mathrm{L}}
\frac{J^2-[\Delta -i\gamma/2][\Delta +i(\gr - \gamma/2)]}{J^2-(\Delta-i\gamma/2)^2},\\
\label{eq:aoutRLambdaOver4}
\mean{\aout^\mathrm{R}}
&=  \mean{\ain^\mathrm{L}} \frac{-|J| \gr}{J^2 - ( \Delta-i\gamma/2)^2}.
\end{align}
As $|J|=\gr/2$, we expect transmission extinction if $\gnr/\gamma \ll 1$ as in the $\lambda/2$ case. However, here the width of the extinction is given by $\gamma$, whereas this width was superradiant $(\Gamma_B=2\gamma)$ in the $\lambda/2$ case.
\begin{figure}[tb] %  figure placement: top, bottom
   \includegraphics[width=\columnwidth]{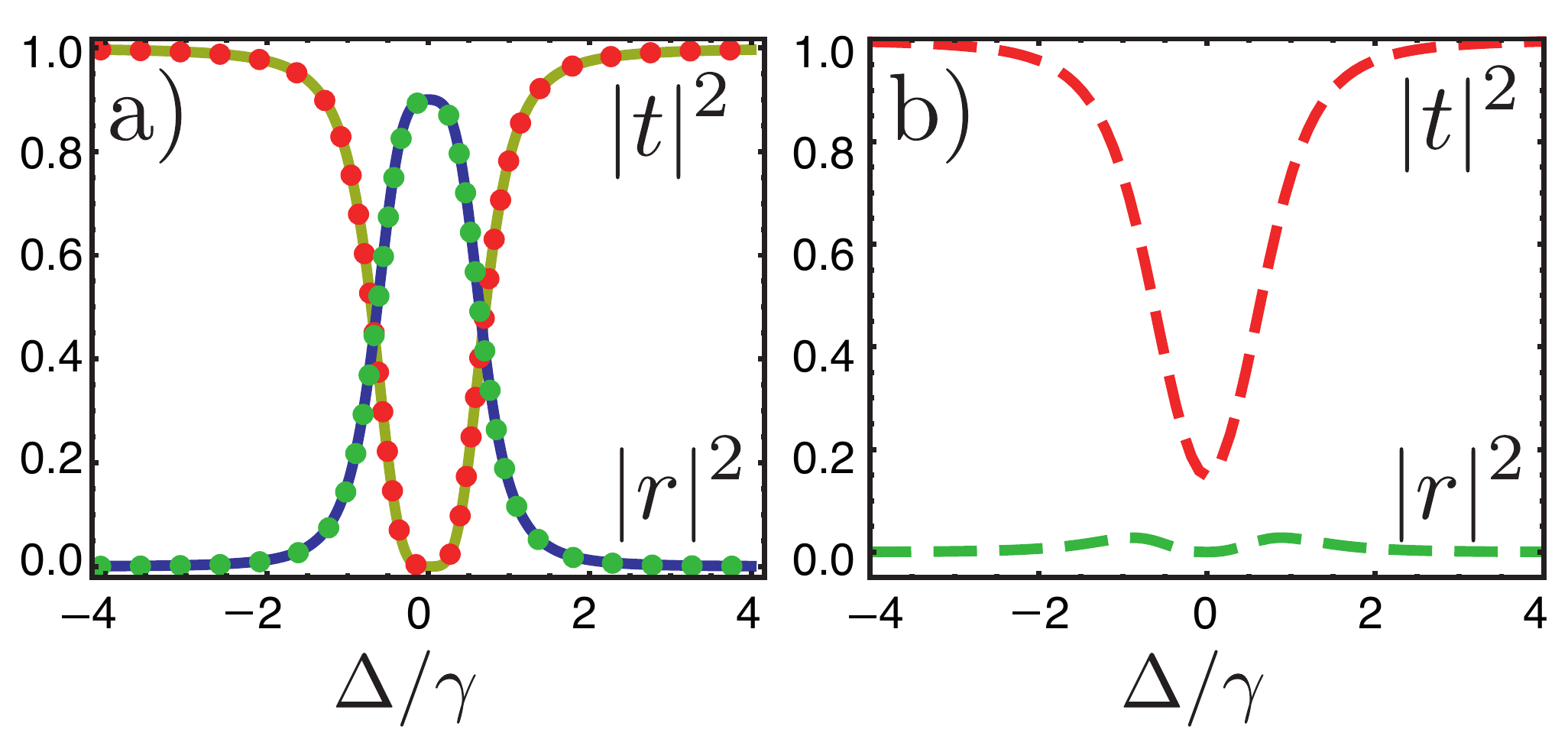} 
   \caption{(Color online) Transmission and reflection coefficients as a function of normalized detuning $\Delta/\gamma$ between the qubits transition frequencies and drive. The qubits are tuned such that $d = 3\lambda_0/4$. Full lines are analytical results while dotted lines are numerical results. (a) $|\epsilon_{0}|/\gamma=0.005$. At low power, $|t|^2+|r|^2 \sim 1$. (b) $|\epsilon_{0}|/\gamma=0.35$. At high power, inelastic scattering is more important so that $|t|^2+|r|^2<1$ around $\Delta=0$. Radiative decay $\gr=0.95 \gamma$ and $\gr/2\pi = 10.3$ MHz in the numerical simulations.}
   \label{fig:aoutLambdaOverFour}
\end{figure}

Using these expressions, we plot in Fig.~\ref{fig:aoutLambdaOverFour}(a) the reflection $|r|^2$ and transmission $|t|^2$ coefficients along with the corresponding results obtained from numerical simulations of the reduced master equation \eq{eq:mainMasterEquation}. The agreement is excellent, with transmission extinction at $\Delta=0$. It is also interesting to observe that these coefficients do not have a Lorentzian profile when nonradiative decay is weak. Indeed, in this situation both $|r|^2$ and $|t|^2$  are rather flat around $\Delta=0$. This is a consequence of the coupling $J$. Since the maximal magnitude of $J$ is $\gr/2$ and the width is $\gamma \ge \gr$, a double peak structure is never resolved and instead leads to the non-Lorentzian profile seen in panel~(a).

In Fig.~\ref{fig:aoutLambdaOverFour}(b), we show results obtained from numerical simulations of the reduced master equation \eq{eq:mainMasterEquation} at a larger power. Because of the increased power broadening, the transmission dip is more Lorentzian-like than in panel~(a). Interestingly, at this higher power we find that $|t|^2+|r|^2 < 1$ for $|\Delta| \lesssim \gamma$. This is because at these powers the effective two-level system becomes strongly dressed by the incoming light, leading to significant inelastic scattering. As expected, in this situation the power spectrum shows a Mollow triplet structure~\cite{Loo:2013,Astafiev:2010,hoi:2012a}. A signature of this dressing can be found in the reflection coefficient which shows two small peaks whose separation is tuned by the input power.  For even larger power, the effective two-level system becomes saturated and $|t|^2 \rightarrow 1$ for all values of $\Delta$ (not shown). This is also observed experimentally \cite{Loo:2013}.

\subsection{Inelastic scattering}

\begin{figure}[tb] %  figure placement: top, bottom
   \includegraphics[width=\columnwidth]{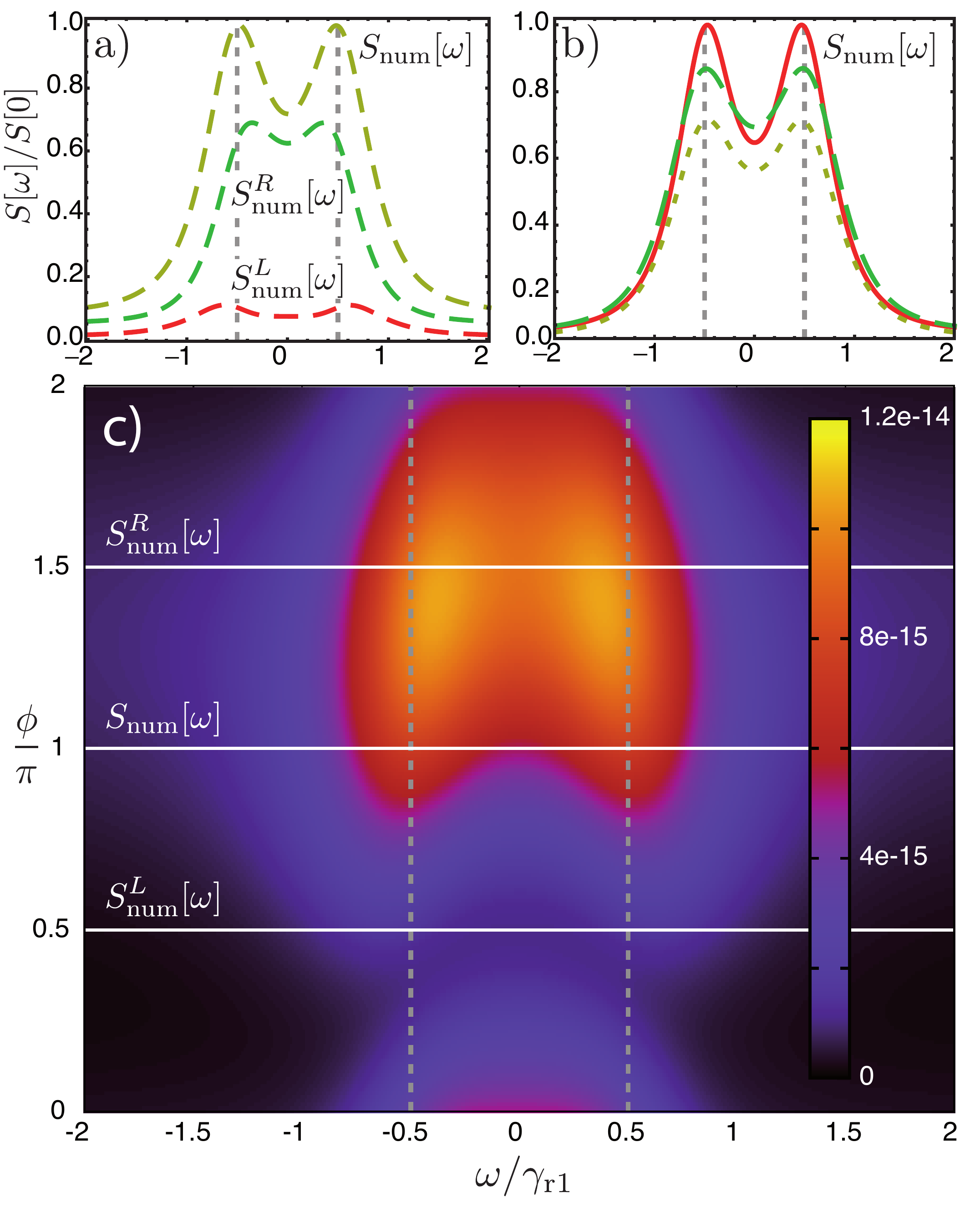} 
   \caption{(Color online) Spectral density as a function of normalized frequency $\omega/\gamma_{\text{r}1}$ where $\gamma_{\text{r}j}=\gamma_{jj}-\gnr$. The qubits are tuned such that $d = 3\lambda_0/4$. The spectral density of the transmitted power $S^\text{L}(\omega)$ shows a signature of the exchange interaction $J$ while it is less obvious in reflection $S^\text{R}(\omega)$. The spectral density of the total signal $S[\omega]$ shows a splitting of $2J$ (indicated by the vertical dashed lines) (a) $\gnr=0$. (b) $(\gamma_{00}-\gamma_{11})/(\gamma_{00}+\gamma_{11})=(0.9,1.1)$ for the solid red line and the long-dashed green line respectively.  $\gnr/\gamma_{\text{r}1}=0.1$ for the dashed yellow line. (c) $S[\omega,\phi]$ as a function of $\omega$ and $\phi$ with $\gnr=0$ and $\gamma_{\text{r}0}=\gamma_{\text{r}1}$. In all cases, $\gamma_{\text{r}1}/2\pi = 10.3$ MHz.}
   \label{fig:lambdaOverFourSpectrum}
\end{figure}

Taking $\gamma=\gamma_{jj}$ as in the previous section, the power spectrum, Eq.~\eqref{eq:DefSomega}, vanishes to second order in $\epsilon_B/\Gamma_D$, where $\epsilon_B$ is the drive amplitude.  In this section, we therefore rely on numerical integration of the master equation \eq{eq:mainMasterEquation}. Figure~\ref{fig:lambdaOverFourSpectrum}(a) shows the simulated power spectrum in transmission $S^\text{L}[\omega]$ (red dashed line) and in reflection $S^\text{R}[\omega]$ (green dashed line) in the absence of nonradiative damping. The yellow dashed line corresponds to the power spectrum $S[\omega]$ obtained by combining the transmitted and reflected components of the output fields with a phase shift of $\pi/2$. In all cases, a clear signature of the exchange interaction $J$ can be seen in the form of a doublet feature in the spectrum. The splitting of this doublet is exactly $2J$ for $S[\omega]$, while it is \emph{larger} than $2J$ when measured only in transmission and \emph{smaller} than $2J$ in reflection. In the presence of nonradiative damping, we thus expect the doublet structure to be easier to resolve in transmission. 
%however with the additional challenge that the overall power there is much smaller than in reflexion and therefore requires more averaging in the presence of amplifier noise. 

To understand the differences between $S^\text{R}[\omega]$, $S^\text{L}[\omega]$, and $S[\omega]$, it is useful to consider the expected spectrum as a function of an additional phase shift. For this reason, we introduce
\be
\label{eq:SpectrumVsPhase}
S[\omega,\phi]= \frac{\gr}{2} \int_{-\infty}^\infty dt e^{i\omega t} \langle A^\dagger (t,\phi) A(0,\phi) \rangle,
\ee
where we have defined
\be
A(t,\phi)=\sigma_-^B(t)+e^{i \phi} \sigma_-^D(t).
\ee
This spectrum is plotted as a function of frequency and phase in Fig.~\ref{fig:lambdaOverFourSpectrum}(c). As noted earlier, the transmitted signal $S^\text{L}[\omega]=S[\omega,\pi/2]$ shows a splitting larger than $2J$, the reflected signal $S^\text{R}[\omega]=S[\omega,3\pi/2]$ a splitting smaller than $2J$, while the splitting of $2J$ is recovered for $S[\omega]=S[\omega,\pi]$. We note that in the above expression for $S[\omega,\phi]$ we have removed the contribution of the in-field for clarity [see Eq.~\eqref{eq:DefSomega} for the full expression].

We explore in Fig.~\ref{fig:lambdaOverFourSpectrum}(b) the effect of asymmetry between relaxation rates and of nonradiative decay on $S[\omega]$.  The only significant contribution is a rescaling of the power spectral density. In the case of asymmetry, this rescaling occurs because the coupling $\gamma_{r0}=\gamma_{00}-\gnr$ between the qubits and the line is changing, whereas in the case of nonradiative decay the rescaling is due to the increased losses.

\section{Conclusion}
\label{sec:conclusion}

Based on Lehmberg's work~\cite{Lehmberg:1970}, we have obtained an effective master equation describing an arbitrary number of inhomogeneous many-level (artificial) atoms coupled to a 1D transmission line and driven by a coherent state. Elastic and inelastic scattering of an input beam are calculated for two qubits using input-output theory.  While individual atoms act as simple mirrors at low power, reflecting incident light, collective effects emerge in the presence of several atoms coupled to the same line. The nature of these effects changes with qubit separation or equivalently with the qubit transition frequency.  When the qubits are separated by $\lambda_0$, elastic and inelastic scattering show signatures of super- and subradiance. The dark state associated with subradiance can be made not completely dark by changing the asymmetry between the qubits' relaxation rates. This can be done by tuning the qubits' transition frequencies. On the other hand, for a separation corresponding to $3\lambda_0/4$, the inelastic scattering shows a doublet structure. This is a signature of the coherent exchange of virtual photons between the atoms. These results are in excellent agreement with experimental results~\cite{Loo:2013}.

Interesting directions for future work include exploring interaction of atoms with a squeezed electromagnetic field in a transmission line~\cite{murch:2013a}, studying correlation function measurements of the transmitted or reflected fields, and considering a network of atoms in a one-dimensional waveguide.

\begin{acknowledgments}
We thank J\'erome Bourassa for useful discussions.  KL and AB acknowledge support from NSERC, CIFAR, and the Alfred P. Sloan Foundation. BCS acknowledges support from AITF, NSERC, and CIFAR. Computations were made on the supercomputer Mammouth parall\`ele II from Universit\'e de Sherbrooke, managed by Calcul Qu\'ebec and Compute Canada. The operation of this supercomputer is funded by the Canada Foundation for Innovation (CFI), NanoQu\'ebec, RMGA and the Fonds de recherche du Qu\'ebec-Nature et technologies (FRQ-NT).
\end{acknowledgments}

\appendix

\section{Derivation of the waveguide QED Hamiltonian with superconducting qubits}
\label{sec:JosephsonJunctions}
\begin{figure}[htbp] %  figure placement: here, top, bottom, or page
    \includegraphics[width=0.85\columnwidth]{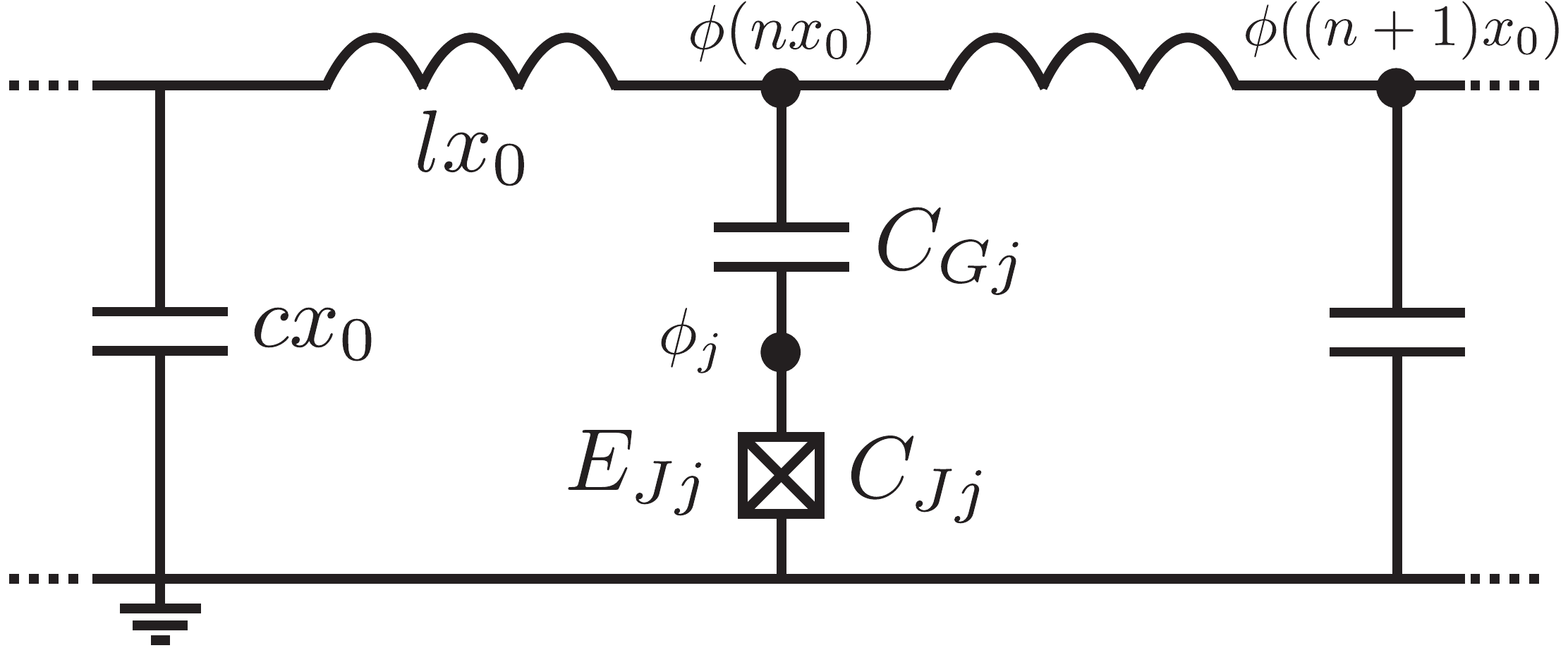} 
   \caption{Lumped element representation of a transmission line, capacitively coupled to a transmon qubit.}
   \label{fig:transmissionLine}
\end{figure}

In this appendix, we derive the Hamiltonian for an ensemble of $N$ transmon qubits capacitively coupled to an open transmission line. A similar calculation for a single qubit can be found in Ref.~\cite{peropadre:2012a}. As illustrated in Fig.~\ref{fig:transmissionLine}, we use a lumped element description of the line, which is characterized by a capacitance per unit length $c$ and inductance per unit length $l$, with $x_0$ the  length of a single $LC$ unit which will be taken to zero below. The $j$th qubit, of Josephson energy $E_{Jj}$ and capacitance $C_{Jj}$, is coupled to the line at positions $x_j$ through a gate capacitor $C_{Gj}$. The corresponding Lagrangian is
\begin{align}
L=&\sum_n \left[\frac{c x_0 }{2} \dot{\phi}(n x_0)^2+ \sum_j \frac{C_{Gj} \delta_{n x_0  , x_j}}{2}[\dot{\phi}(n x_0)-\dot{\phi}_j]^2\right.\nnn
&\left.-\frac{\left\{\phi[(n+1)x_0]-\phi(n x_0)\right\}^2}{2l x_0}\right]  \nnn
&+\sum_{j=0}^{N-1} \left[ \frac{C_{Jj}}{2}\dot{\phi}_j^2 + E_{Jj} \cos\left( \frac{2\pi}{\Phi_0} \phi_j  \right) \right],
\end{align}
where $\phi(x)$ and $\phi_j$ are the generalized fluxes, as defined in Ref.~\cite{devoret:1995a}. $\Phi_0$ is the magnetic flux quantum. Introducing the charges $p(x_j )$ and $p_j$ conjugate to the generalized fluxes, the Hamiltonian takes the form
\begin{align}
 H_\mathrm T=&H_\mathrm A+H_\mathrm F+\sum_j \frac{p_j p(x_j )+\hc}{2c_{gj}}.
 \end{align}
In this expression,
\be
H_\mathrm A=\sum_j \left[\frac{p_j^2}{2C_j}  - E_{Jj} \cos \left( \frac{2\pi}{\Phi_0} \phi_j  \right) \right] 
\ee
is the free-transmon Hamiltonian with 
\be
C_j=\frac{ (C_{Gj}+C_{Jj})c x_0 +  C_{Gj}C_{Jj}}{C_{Gj}+c x_0}.
\ee
The transmission line Hamiltonian reads 
\be
H_\mathrm F=\sum_n \left(\frac{x_0 p(n x_0 )^2}{2c_L(n x_0 )} + \frac{\left\{\phi[(n+1) x_0]-\phi(n x_0)\right\}^2}{2l x_0}\right),
\ee
where we have defined the effective transmission line capacitance per unit of length
\be
c_L(n x_0) = c +\sum_j \frac{C_{Gj}C_{Jj}}{C_{Gj}+C_{Jj}}\frac{\delta_{n x_0,x_j}}{x_0},
\ee
and the effective coupling capacitance per unit of length
\be
c_{gj}=\frac{\left( C_{Gj}+C_{Jj}\right) c + C_{Gj}C_{Jj}/x_0}{C_{Gj}}.
\ee

Letting $x_0 \rightarrow 0$, we obtain
\be
H_\mathrm F=\int dx \left(\frac{p(x)^2}{2c_L(x)}+ \frac{\left[\partial_x \phi(x)\right]^2}{2l}\right),
\ee
with
\begin{align}
C_j=&\frac{ (C_{Gj}+C_{Jj})c L_A +  C_{Gj}C_{Jj}}{C_{Gj}+c L_A},\\
c_L(x) =& c +\sum_j \frac{C_{Gj}C_{Jj}}{C_{Gj}+C_{Jj}}\delta(x-x_j),\\
\label{eq:couplingCapacity} c_{gj}=&\frac{\left( C_{Gj}+C_{Jj}\right) c + C_{Gj}C_{Jj}/L_\text{A}}{C_{Gj}},
\end{align}
with $L_\text{A}$ the length of the Josephson junctions.

Using the expression for $c_L(x)$ above, the average capacitance of the transmission line over a length $d$ extending over all of the qubits is
\begin{align}
\bar c_L  &= \frac{1}{d}\int_0^d dx \left( c+\sum_j \frac{C_{Gj} C_{Jj}}{C_{Gj}+C_{Jj}} \delta(x-x_j) \right)\nnn
&=c+\frac{1}{d}\sum_j \frac{C_{Gj} C_{Jj}}{C_{Gj}+C_{Jj}}.
\end{align}
In practice, $cd \gg \sum_j C_{Gj} C_{Jj}/(C_{Gj}+C_{Jj})$ and the qubit's capacitances are small perturbations on the transmission line. In other words, $c_L(x) \sim c$ and the standard quantization procedure leads to the Hamiltonians of Eqs.~\eq{eq:fullHfield} - \eq{eq:fullHinteraction} in the main text \cite{Romero:2009}.

\section{Master equation for ensemble of inhomogeneous atoms in open 1D space}
\label{sec:masterEquation}

For completeness, we derive in this appendix the master equation presented in Eq.~\eqref{eq:mainMasterEquation} of the main text. The different approximations used in obtaining this result are mentioned here, but discussed in more detail in Appendix~\ref{sec:discussionApproximations}. As discussed in Sec.~\ref{sec:WaveguideQED}, we consider an ensemble of $N$ multi-level (artificial) atoms dipole coupled to a 1D transmission line. The total Hamiltonian takes the form $H_\mathrm{T}=H_\mathrm{F}+H_\mathrm{A}+H_\mathrm{I}$, where the field Hamiltonian $H_\mathrm{F}$ is given in Eq.~\eqref{eq:fullHfield}, the artificial atom Hamiltonian $H_\mathrm{A}$ in  Eq.~\eqref{eq:fullHatoms}, and their interaction $H_\mathrm{I}$ in Eq.~\eqref{eq:fullHinteraction}.

Following Lehmberg~\cite{Lehmberg:1970}, to obtain an effective master equation for the artificial atoms we first move to the Heisenberg picture where the field operator $\dot{a}_\mathrm{R}(\omega)$ obeys the equation
\begin{align}\label{eq:dot_a_R}
\dot{a}_\mathrm{R}(\omega)&=-i\omega a_\mathrm{R}(\omega)+ \sum_{mj} g_j \sqrt{m+1} \sqrt{\omega}  e^{-i \omega x_j/v}\sigma_x^{mj}.
\end{align}
Integrating from an initial time $t_0=0$ before the interaction, the above equation yields
\begin{align}
\label{eq:a_solution_ini}
a_\mathrm{R} (\omega,t)=&a_\mathrm{R}(\omega,0) e^{-i \omega t}\nnn
&+\sum_{mj} g_j  \sqrt{m+1}  \sqrt{\omega} \int_0^t d\tau e^{-i \omega(t-\tau+t_j)} \sigma_x^{mj}(\tau),
\end{align}
with $t_j=x_j/v$. The expression for $a_\mathrm{L}(\omega,t)$ is obtained with the replacement $t_j \rightarrow -t_j$. Using these results we can express $\Xi_j(t)$ defined in Eq.~\eqref{eq:define_xi} as
\begin{align}
\label{eq:exactXi}
\Xi_j(t)
=&  \Xi^\mathrm{in}_j(t) \nnn 
&-i\sum_{nk}\sum_{\sigma=\pm 1} g_k \sqrt{n+1} \int_0^t d\tau I_{nk}(t,\tau,\sigma t_{kj}),
\end{align}
with $t_{kj}= |x_k - x_j|/v$ the time a signal takes to propagate from atom $k$ to atom $j$ and where we have defined
\begin{equation}\label{eq:Xi_in}
\begin{split}
\Xi^\mathrm{in}_j(t)
=-i\int_0^\infty d\omega \sqrt{\omega}
&\left[a_\mathrm{L}(\omega,0) e^{-i \omega (t+t_j)}\right.\\
&\left.+a_\mathrm{R}(\omega,0) e^{-i \omega (t-t_j)}\right]
\end{split}
\end{equation}
and
\begin{align}
\label{eq:ink}
I_{nk}(t,\tau,t_{kj}) =  \int_0^\infty d\omega   \omega e^{i \omega(\tau-t- t_{kj})} \sigma_x^{nk}(\tau).
\end{align}
Since the integrand of $I_{nk}(t,\tau,t_{kj})$ is proportional to $\omega$, the integral is dominated by high frequencies where the exponential is, however, oscillating rapidly. As argued in Appendix~\ref{sec:discussionApproximations}, it is then reasonable to take 
\be
\label{eq:approxMarkov}
\sigma_-^{nk}(\tau)
\approx \sigma_-^{nk}(t)e^{-i\omega_{nk}(\tau-t)},
\ee
and $\omega_{nk} t \rightarrow \infty$, where
\be
\omega_{nk} = \left(E_{(n+1)k}- E_{nk}\right)/\hbar
\ee
is the transition frequency between levels $n+1$ and $n$ of atom $k$.

Using the standard identity
\be
\label{eq:usefulIdentity}
\int^{\infty}_0 dxe^{-i kx}=\pi \delta \left( k \right)-i \text{P}\left(\frac{1}{k}\right),
\ee
with $\text{P}$ the Cauchy principal value, we obtain
\begin{align}
\Xi_j(t)=&\Xi^\mathrm{in}_j(t)\nnn
&-\frac{1}{g_j}\sum_{nk}    \left[ \Omega^{n+}_{kj} \sigma_+^{nk}+\left( \Omega^{n-}_{kj}+i\gamma_{kj}^n/2 \right)  \sigma_-^{nk} \right].
\end{align}
In this expression, we have defined 
\begin{equation}
\label{eq:coeffOmegaDefinition}
\Omega^{n\pm}_{kj} = 2 g_k g_j \sqrt{n+1}\, \text{P}\int_0^\infty     \frac{\omega \cos\left[\omega t_{kj}\right]}{\omega\pm\omega_{nk}}d \omega
\end{equation}
and
\begin{equation}
\label{eq:coeffGammaDefinition}
\gamma_{kj}^n =4 \pi g_k g_j  \omega_{nk} \sqrt{n+1}    \cos\left[\omega_{nk} t_{kj}\right].
\end{equation}

Again following Lehmberg~\cite{Lehmberg:1970}, a reduced master equation for the atoms is obtained by first considering the Heisenberg equation of motion of an arbitrary operator $Q$ acting on the atoms only. Given that $[Q(t),\Xi_j(t)]=0$ at all times [this is more clearly seen from the form of Eq.~\eqref{eq:define_xi} of $\Xi_j(t)$], we find
\begin{align}
\label{eq:operatoronqubits}
\dot{Q}(t) =&\frac{i}{\hbar}\left[H_A+\hbar \sum_{mj} \sqrt{m+1}g_j \left(\Xi^\mathrm{in}_j+\hc\right)\sigma_x^{mj},Q\right]\nnn
&+\sum_{mj}\sum_{nk}\sqrt{m+1}\nnn
& \times \bigg[  -i\Omega^{n+}_{kj} \left(  \sigma_x^{mj}Q\sigma_+^{nk}-Q\sigma_x^{mj}\sigma_+^{nk}-\hc\right) \nnn
&  -i \Omega^{n-}_{kj}\left(\sigma_x^{mj}Q\sigma_-^{nk} -Q\sigma_x^{mj} \sigma_-^{nk}-\hc\right)\nnn
&+  \frac{\gamma_{kj}^n}{2}\left(\sigma_x^{mj}Q \sigma_-^{nk}-Q\sigma_x^{mj} \sigma_-^{nk}+\hc  \right)\bigg].
\end{align}
With $\mathrm{Tr}(\dot{Q} \rho_T) =\mathrm{Tr_A}(Q \dot{\rho})$, where $\rho=\mathrm{Tr_\mathrm{R}}(\rho_T)$ is the reduced master equation of the atoms, we finally obtain after using the rotating-wave approximation
\begin{equation}
\begin{split}
\label{eq:lastExactMasterEquation}
\dot{\rho}=&-i\left[\frac{H_\text{A}}{\hbar} +\sum_{mj}L_{mj} \ket{m_j}\bra{m_j},\rho\right]\\
&-i\left[\sum_{mj} d_{mj}(t)\sigma_x^{mj}+\sum_{mj}\sum_{nk} J_{mj,nk}\sigma_+^{nk} \sigma_-^{mj},\rho\right]\\
&+\sum_{mj}\sum_{nk}\bigg[ \gamma_{mj,nk} \left(\sigma_-^{mj}\rho\sigma_+^{nk} -\frac{1}{2}\left\{\sigma_+^{nk} \sigma_-^{mj},\rho\right\}\right)\\
&+\Omega_{mj,nk} \left(  \sigma_+^{mj}\rho\sigma_-^{nk}+\sigma_-^{nk}\rho\sigma_+^{mj} -\left\{\sigma_-^{nk}\sigma_+^{mj},\rho\right\} \right)\bigg].
\end{split}
\end{equation}

To obtain this expression, we have assumed that the system is driven from the left and the right by coherent fields of frequency $\omega_d$, phase $\theta_{\text{L(R)}}$ and power $P_{\text{L(R)}}$. As shown in Appendix~\ref{sec:drive}, the resulting amplitude of the qubit driving term in the above master equation takes the form
\begin{align}\label{eq:drive_amplitude_full}
d_{mj}(t)=&-2\sqrt{\frac{\gamma_{mj,mj}}{2}} \left( \sqrt{\frac{P_\mathrm{L}}{\hbar\omega_{mj}}}\sin\left[ \omega_d (t+t_j+\theta_\mathrm{L})\right] \right.\nnn
&\left.+\sqrt{\frac{P_\mathrm{R}}{\hbar\omega_{mj}}}\sin \left[\omega_d (t-t_j+\theta_\mathrm{R}) \right]\right).
\end{align}
We have also defined the Lamb shift
\begin{align}
L_{mj}=& -\left(\sqrt{m+1}\Omega^{m+}_{jj}-\sqrt{m}\Omega^{(m-1)-}_{jj}\right),
\end{align}
the (joint) decay rate
\begin{align}
\label{eq:ApprelaxationRate}
\frac{\gamma_{mj,nk}}{2 \pi}=& g_k g_j \sqrt{(m+1)(n+1)} 
 \left(\chi_{mkj}+\chi_{nkj}^*\right),
\end{align}
with $\chi_{mjk}=\omega_{mj} e^{i \omega_{mj} t_{kj}}$, the atom-atom exchange interaction amplitude
\begin{align}
\label{eq:AppexchangeInteraction}
\frac{J_{mj,nk}}{2 \pi}=& -i \frac{g_k g_j}{2} \sqrt{(m+1)(n+1)}\left( \chi_{njk}-\chi_{mkj}^*\right),
\end{align}
and
\be
\Omega^{}_{mj,nk}=-i\left(\sqrt{n+1}\Omega^{m+}_{jk} - \sqrt{m+1}\Omega^{n+}_{kj}\right).
\ee
The reader interested in (even) more details will find the evaluation of the integrals needed to find these coefficients  in Appendix~\ref{sec:coefficientEvaluation}.

In the main text, we absorb the Lamb shift $L_{mj}$ into the definition of the atomic transition frequency. As usual for point like atoms, this contribution is formally infinite. A nondiverging result can be obtained by taking into account the finite size of the atoms~\cite{bourassa:2013a}. In any case, experimentally, the Lamb shift is always present in the evaluation of the various transition frequencies and absorbing it in the definition here does not cause any problems.  

We also note that the matrix of components $\Omega_{mj,nk}$ is traceless, Hermitian, and nonzero. As a result, it is not semipositive as is required to express the master equation in Lindbladian form. Fortunately, and as discussed in Appendix~\ref{sec:discussionApproximations}, the various $\Omega_{mj,nk}$ are in practice small and can safely be neglected. Doing so, we finally arrive at the the master equation, Eq.~\eqref{eq:mainMasterEquation}, with an effective Hamiltonian given in Eq.~\eqref{eq:H}.

\section{Discussion of the approximations}
\label{sec:discussionApproximations}

In this appendix, we discuss in more detail the main approximations that have been used to obtain the reduced master equation. These approximations are certainly not original to this work and this discussion is added for completeness.

\subsection{Markov approximation}

We first start with $I_{nk}(t,\tau,t_{kj})$ defined in Eq.~\eq{eq:ink}. The complexity in this expression is the dependence of the integrand $\sigma_x^{nk}$ on the integration variable $\tau$. This can be simplified by rewriting $I_{nk}(t,\tau,t_{kj})$ as
\be\begin{split}
\label{eq:markovApproximationNeeded}
I_{nk}(t,\tau,t_{kj})
&=  \int_0^\infty d\omega   \omega e^{i \omega(\tau-t- t_{kj})} \sigma_x^{nk}(\tau)\\
&=  \int_0^\infty d\omega   \omega e^{-i \omega(\tau-t-t_{kj})} \nnn
&\times 
\left[
e^{i H_\text{T} (\tau-t) / \hbar}\sigma_-^{nk}(t) e^{-i H_\text{T} (\tau-t)/\hbar} +\hc
\right].
\end{split}\ee
with $H_\text T=H_\text A+H_\text I+H_\text F$. Since the integrand is proportional to $\omega$, high frequencies contribute most. However, at high frequency, the exponential is oscillating rapidly and the contribution averages out to zero. The only situation where the exponential does not oscillate is when $\tau = t+t_{kj}$ and this is the only value of $\tau$ that we retain.

To simplify this expression further, we neglect the interaction Hamiltonian $H_\mathrm I$ compared to the free Hamiltonian $H_\mathrm A$. More formally, the error $\mathcal E$ that is made in neglecting the interaction goes as~\cite{Poulin:2011}
\begin{align}
\mathcal  E \sim \average{  \frac{ H_\text I ^2}{ H_\text A }}\frac{t_{kj}}{h}.
\end{align}
Evaluating the expectation value of the field operators appearing in $H_\mathrm I$ by assuming a coherent drive of power $P$, this error can be expressed as
\begin{align}
\label{eq:errorNeglectingHi}
\mathcal E \sim&  \frac{32 \pi g_k^2  L}{v\omega_{nk}}\frac{P}{h},
\end{align}
with $L$ the maximum distance between two atoms. 

For waveguide QED with transmon qubits, we find 
\be
g_k=\left(\sqrt{\frac{ e^2 c}{2\hbar \pi v c^2_{gk}}} \right) \left(\frac{E_{Jk}}{8E_{Ck}} \right)^{(1/4)},
\ee
with $E_{Jk}$ and $E_{Ck}$ the Josephson and charging energy of the $k$th transmon \cite{Koch:2007}, $c$ the capacitance per unit length of the transmission line, and where $c_{gk}$ is defined in Eq.~\eqref{eq:couplingCapacity}. Using typical experimental values for these parameters~\cite{Fink:2010a}, we find that $g_k \sim 0.02$. The numerical value of $g_k$ can also be estimated from the experimentally measured value of the relaxation rate $\gamma_{nk,nk}$. Doing so using recent experimental results~\cite{Loo:2013,Hoi:2012,Astafiev:2010} gives consistent results. Now, given that the speed of light in the transmission line is $v\sim 10^8$ m/s and assuming a separation $L \sim 1$ cm between two transmons of transition frequency $\omega_{nk}\sim 2\pi \times 6$ GHz, we find $\mathcal E\sim0.02$ for $P\sim -100$~dBm. Since this power is large in practice~\cite{Loo:2013}, dropping the contribution of $H_\mathrm I$ from Eq.~\eqref{eq:markovApproximationNeeded} is reasonable. Doing so we can rewrite $\sigma_-^{nk}(\tau)$ in $I_{nk}(t,\tau,t_{kj})$ as
\be
\sigma_-^{nk}(\tau) \approx \sigma_-^{nk}(t)e^{-i\omega_{nk}(\tau-t)}.
\ee
This corresponds to a Markov approximation. It breaks down for larger separation, i.e., for $\omega_{nk} t_{kj}/(2\pi) \sim 10$, something that was studied in Ref.~\cite{Zheng:2012a}.

It is interesting to note that, even if we recover a result similar to Lehmberg's~\cite{Lehmberg:1970}, here we used a different justification. Indeed, Lehmberg assumed that the atoms are close enough such that the time it takes for a signal to propagate from one atom to the other is small compared to the Larmor frequencies, $L \ll v/\omega_{nk}$. This is inapplicable in a waveguide QED setup with superconducting qubits.

\subsection{Long-time approximation and causality}

The identity~\eqref{eq:usefulIdentity} is essential in deriving the master equation~\eqref{eq:mainMasterEquation}. To use this identity, we need the upper bound of the time integral in Eq.~\eqref{eq:exactXi} to go to infinity. Using approximation \eq{eq:approxMarkov},
\begin{align}
\int_0^t d\tau I_{nk}(t,\tau,t_{kj})=&  \int_0^t d\tau \int_0^\infty d\omega   \omega e^{-i \omega(\tau-t-t_{kj})} \nnn
&\times \left(\sigma_-^{nk} e^{-i\omega_{nk}(\tau-t)}+\hc\right),
\end{align}
and with a change of variable $x= \omega_{nk} (t-\tau)$,
\begin{align}
\int_0^t d\tau& I_{nk}(t,\tau,t_{kj})= \int_0^\infty d\omega  \int^{\omega_{nk} t}_0 dx \frac{\omega e^{-i \omega t_{kj}}}{\omega_{nk}}  \nnn
&\times \left[ \sigma_-^{nk}e^{-i(\omega-\omega_{nk})x/\omega_{nk}}  +\sigma_+^{nk}e^{-i (\omega_{nk}+\omega)x/\omega_{nk}}    \right].
\end{align}
The integrand of $x$ is an oscillating function. If $\omega_{nk}t \gg 1$, the integration is already over many periods of this function, so it is a good approximation to let $\omega_{nk}t \rightarrow \infty$. For waveguide QED with superconducting qubits, this condition requires that $t \gg 0.02$ ns. Since we are not interested in dynamics at this very fast time scale, this approximation holds here. In fact, the electronics in typical experiments have a bandwidth of less than $\sim1$ GHz~\cite{Fink:2010a}. The same argument justifies taking $\omega_{mj}(t-t_f)\rightarrow - \infty$.

With this approximation, the atoms are treated as interacting instantaneously. This is not a major problem because the phase shift associated with the delays it takes for light to travel from one atom to another is taken into account by the factor $\exp(-i\omega t_{kj})$. Hence, interaction between two atoms at time $t$ is mediated through light that has been emitted by the atoms at an earlier time $t-t_{kj}\sim t-L/v$. A more important problem arises during transients. For example, assume a drive is suddenly turned on such that it affects a first atom at time $t$. Our model causes the drive to affect the second atom at this very time $t$ with a phase delay $\exp(-i\omega t_{kj})$ rather than at a time $t+t_{kj}$. As a result, we do not expect transient effects on a time scale smaller than $L/v$ to be correctly captured. This is not an issue for the steady-state quantities that are computed here and measured in Ref.~\cite{Loo:2013}.

\subsection{RWA, infinite terms, and Lindblad form}

In this section, we justify the approximations that were made in going from the Heisenberg equation of motion~\eqref{eq:operatoronqubits}  to the master equation~\eq{eq:mainMasterEquation}. The error made by making the rotating-wave approximation can be expressed as~\cite{Kliesch:2011} 
\be
\mathcal E \sim \frac{8M^2\left[ \left(\gamma_{kj}^n\right)^2 + \left(\Omega^{n+}_{kj}\right)^2+\left(\Omega^{n-}_{kj}\right)^2\right]}{3(\omega_{mj}+\omega_{nk})^2}.
\ee
This error is formally infinitely large simply because $\Omega^{n\pm}_{jj} \rightarrow \infty$ as shown in Appendix~\ref{sec:coefficientEvaluation}. This divergence is present because we did not take into account the physical dimensions of the artificial atoms (either explicitly or with a cut-off frequency). Ignoring this unphysical problem and using typical circuit QED parameters yields $\mathcal E \sim 0.001$. Considering this, we can safely make the rotating-wave approximation.

To go from Eq.~\eq{eq:lastExactMasterEquation} to its Lindblad form, Eq.~\eqref{eq:mainMasterEquation}, we also need to neglect terms proportional to $\Omega_{mj,nk}$ since this matrix is not semipositive. These terms are either small or quickly rotating. The error made by dropping them is, considering that the relevant time scale goes as $\gamma^{-1}_{mj,mj}$~\cite{Kliesch:2011},
\be
\mathcal E \sim \frac{8M^2|\Omega_{mj,nk}|^2}{3(\gamma_{mj,mj})^2}
\ee
so that $\mathcal E \sim 0.002$. Hence, once more, it is a very good approximation to neglect these terms.

\section{Driving term}
\label{sec:drive}

To proceed from Eq.~\eq{eq:operatoronqubits} to~\eq{eq:lastExactMasterEquation}, we need to take care of the terms proportional to $\Xi^\mathrm{in}_j$ and ${\Xi^\mathrm{in}_j}^\dagger$ since these operators contain contributions from the field operator $\ain^{\text{R/L}}$. We regroup these terms under what we call the drive superoperator $D$ acting on $\rho$ in the master equation such that
\begin{align}
\trs{Q D\rho}{\mathrm A}=&i\sum_{mj} g_j \sqrt{m+1}\nnn
& \times \trs{\trs{ \left( \left[\sigma_x^{mj},Q\right] \Xi_j^\mathrm{in}-\hc\right) \rho_\text T}{\text R}}{\text A}\nnn
=&\left\langle i\sum_{mj} g_j \sqrt{m+1} \left( \left[\sigma_x^{mj},Q\right] \Xi_j^\mathrm{in}-\hc\right) \right\rangle.
\end{align}
Here, all operators are evaluated at time $t$. Because of causality, the input field operator $\Xi_j^\mathrm{in}$ cannot be correlated with any atomic operator when they are both evaluated at the same time. Therefore,
\begin{align}
\trs{Q D\rho}{\text A}=&\sum_{mj}  \left\langle ig_j \sqrt{m+1} \left( \left[\sigma_x^{mj},Q\right] \Xi_j^\mathrm{in}-\hc\right) \right\rangle\nnn
=& i\sum_{mj} g_j \sqrt{m+1} \left(\left\langle \left[\sigma_x^{mj},Q\right] \right\rangle \left\langle\Xi_j^\mathrm{in} \right\rangle -\hc\right) \nnn
=&i\sum_{mj} g_j \sqrt{m+1}\nnn
&\times \trs{ \left[ \left( \left\langle\Xi_j^\mathrm{in} \right\rangle  + \text{c.c.} \right)\sigma_x^{mj},Q\right] \rho  }{\text A},
\end{align}
which leads to
\begin{align}
\dot{\rho}=&-i\left[\frac{H_\text A}{\hbar}+\sum_{mj}d_{mj}(t)\sigma_x^{mj},\rho\right]+\sum_{mj}\sum_{nk} \sqrt{m+1} \nnn
& \times \left[ -i\Omega^{n+}_{kj} \left(  \sigma_+^{nk}\rho\sigma_x^{mj} -\sigma_x^{mj}\sigma_+^{nk}\rho-\hc\right) \right.\nnn
& -i \Omega^{n-}_{kj}\left(\sigma_-^{nk}\rho\sigma_x^{mj}-\sigma_x^{mj} \sigma_-^{nk}\rho-\hc\right)\nnn
&+\left.   \frac{\gamma_{kj}^n}{2}\left(\sigma_-^{nk}\rho\sigma_x^{mj} -\sigma_x^{mj} \sigma_-^{nk}\rho -\hc\right)\right],
\end{align}
with
\be
d_{mj}(t)=g_j \sqrt{m+1} \left( \left\langle\Xi_j^\mathrm{in}(t) \right\rangle  +  \left\langle\Xi_j^\mathrm{in}(t) \right\rangle^* \right).
\ee
We make the assumption that we are driving at frequency $\omega_d$ with a coherent state $\ket{\{\alpha\}}$ such that \cite{Loudon:2007}
\be
a_\mathrm{L (R)} (\omega,0) \ket{\{\alpha\}}=\sqrt{\frac{P_{\mathrm{L (R)}}}{\hbar \omega_d} }e^{-i \omega_d \theta_{\mathrm{L (R)}}} \delta(\omega-\omega_d)\ket{\{\alpha\}},
\ee
with $P_{\text{L(R)}}$ and $\theta_{\text{L(R)}}$, respectively, the power and phase of left (right) movers. Using this, we have
\begin{align}
\left\langle\Xi_j^\mathrm{in}(t) \right\rangle =&\bra{\{\alpha\}}\Xi^\mathrm{in}_j\ket{\{ \alpha \}}\nnn
=&-i\left[e^{-i \omega_d (t+t_j+\theta_\mathrm{L})} \sqrt{2 \pi P_\mathrm{L}/\hbar}\right.\nnn
&\left. + e^{-i \omega_d (t-t_j+\theta_\mathrm{R})} \sqrt{2 \pi P_\mathrm{R}/\hbar} \right].
\end{align}
Finally, the drive rate can be written as
\begin{align}
d_{mj}(t)=&-2\sqrt{\frac{\gamma_{mj,mj}}{2}} \left( \sqrt{\frac{P_\mathrm{L}}{\hbar\omega_{mj}}}\sin\left[ \omega_d (t+t_j+\theta_\mathrm{L})\right] \right.\nnn
&\left.+\sqrt{\frac{P_\mathrm{R}}{\hbar\omega_{mj}}}\sin \left[\omega_d (t-t_j+\theta_\mathrm{R}) \right]\right).
\end{align}

\section{Evaluation of $\boldsymbol{\Omega^{n\pm}_{kj}}$}
\label{sec:coefficientEvaluation}

In this section, we present details of the integration of  Eq.~\eq{eq:coeffOmegaDefinition}. With the change of variables $x=(\omega \pm \omega_{nk})/\omega_{nk}$ and $y = x \mp 1$, Eq.~\eqref{eq:coeffOmegaDefinition} can be expressed as
\begin{align}
\Omega^{n\pm}_{kj}=&2 g_k g_j \omega_{nk} \sqrt{n+1} \left(\int_{0}^\infty    d y  \cos\left(\omega_{nk} t_{kj} y \right) \right. \nnn
&\left. \mp \text{P}\int_{\pm 1}^\infty  d x   \frac{\cos\left[\omega_{nk} t_{kj} (x \mp 1) \right]}{x}\right).
\end{align}
To deal with the first term of the right-hand-side, we add a converging factor. This reflects the fact that  the system stops to respond at infinite frequencies. In this way, we find that this first term vanishes
\be\begin{split}
\lim_{\eta \rightarrow 0^+}  \int_{0}^\infty d y &\cos\left(\omega_{nk} t_{kj} y \right) e^{-\eta y}\\
=&\lim_{\eta \rightarrow 0^+} \frac{\eta}{(\omega_{nk} t_{kj})^2+\eta^2} = 0.
\end{split}\ee
On the other hand, for the second term, which we denote $I^{\pm}$, we find 
\be\begin{split}
I^{\pm}=&\text{P}\int_{\pm 1}^\infty   dx  \frac{\cos\left[\omega_{nk} t_{kj} (x \mp 1) \right]}{x}\\
=&\cos\left(\omega_{nk} t_{kj}  \right)\text{P}\int_{\pm 1}^\infty     dx \frac{\cos\left(\omega_{nk} t_{kj} x \right)}{x}\\
& \pm\sin\left(\omega_{nk} t_{kj}  \right) \text{P}\int_{\pm 1}^\infty   dx  \frac{\sin\left(\omega_{nk} t_{kj} x \right)}{x}\\
=&-\cos\left( \omega_{nk} t_{kj} \right) \cosInt{ |\omega_{nk} t_{kj}| }\\
&+\frac{\sin\left( \omega_{nk} t_{kj} \right)}{2}\left(\pm\pi \sign{\omega_{nk} t_{kj}} - 2 \sinInt{\omega_{nk} t_{kj} } \right),
\end{split}\ee
where $\cosInt{x}$ and $\sinInt{x}$ are the cosine and sine integral functions:
\be
\cosInt{x}=-\int_x^\infty dt \frac{\cos  t}{t},
\quad
\sinInt{x}=\int_0^x dt \frac{\sin  t}{t}.
\ee
Using these results, we finally obtain 
\begin{align}\label{eq:OmegaPlus}
\Omega^{n\pm}_{kj}=& 2 \pi g_k g_j \omega_{nk} \sqrt{n+1} \nnn
&\times \left[\mp p(\omega_{nk} t_{kj}) +\sin\left( \omega_{nk} t_{kj} \right)\left(\frac{\pm 1-1}{2}\right)\right],
\end{align}
where we have defined a ``proximity'' function
\be\label{eq:ProximityFunction}
p(x)\equiv\frac{\sin\left(|x| \right)\left[ \pi-2\sinInt { |x|}\right]-2\cos\left( x \right) \cosInt{ |x|}}{2\pi}.
\ee
This choice of name reflects the fact $p(x)$ goes to $\infty$ as $x \rightarrow 0$, and rapidly approaches $0$ as $x \rightarrow 1$.
\section{Input-output theory}
\label{sec:inputOutput}
In this section, we derive the input-output boundary condition in the presence of the artificial atoms in the line. This will allow us to compare the theoretical predictions to experiments measuring reflection and transmission.  To derive the reduced master equation, we used the formal solution to the Heisenberg equation of motion for $a_\mathrm{R/L}(\omega,t)$. In Eq.~\eqref{eq:a_solution_ini}, this solution was given for the case where the equation of motion is integrated starting from a time $t_0 = 0<t$ before the interaction. Following the standard input-output prescription~\cite{collett:1984a}, it is also useful to obtain this solution by integrating up to a time $t_f>t$ after the interaction:
\begin{align}
a_\mathrm{R} &(\omega,t)=a_\mathrm{R}(\omega,t_f) e^{-i \omega t}\nnn
&-\sum_{mj} g_j  \sqrt{m+1}  \sqrt{\omega} \int_t^{t_f} d\tau e^{-i \omega(t-\tau+x_j/v)} \sigma_x^{mj}(\tau).
\end{align}
Adding this expression to Eq.~\eq{eq:a_solution_ini} and integrating over $\omega$, we arrive at the input-output boundary condition
\begin{align}
a_{\text{out}}^\mathrm{R}(t)=&a_{\text{in}}^\mathrm{R}(t)+\sum_{mj} g_j  \sqrt{m+1} \nnn
&\times \int_0^\infty \frac{d\omega}{\sqrt{2\pi}} \sqrt{\omega} \int_0^{t_f} d\tau e^{-i \omega(t-\tau+x_j/v)} \sigma_x^{mj}(\tau),
\end{align}
where, similarly to Eq.~\eq{eq:Xi_in}, we have defined the input field
\be
a_{\text{in}}^\mathrm{R}(t)=\int_0^\infty \frac{d\omega}{\sqrt{2\pi}} a_\mathrm{R}(\omega,0) e^{-i \omega t},
\ee 
and the output field
\be
a_{\text{out}}^\mathrm{R}(t)=\int_0^\infty \frac{d\omega}{\sqrt{2\pi}} a_\mathrm{R}(\omega,t_f) e^{-i \omega t}.
\ee
While $a_{\text{in}}^\mathrm{R}(t)$ can be interpreted as the field incident on the system from the left, $a_{\text{out}}^\mathrm{R}(t)$ represents the field propagating to the right after interaction with the system.

It is possible to express the boundary condition in a more useful form by using the approximation of Eq.~\eq{eq:approxMarkov}. Indeed, making the change of variable $y= \omega_{mj} (t-\tau)$ and taking $\omega_{mj}(t-t_f)\rightarrow - \infty$, we find the simpler form
\begin{align}
a_{\text{out}}^\mathrm{R}(t)&=a_{\text{in}}^\mathrm{R}(t)+\sum_{mj}  e^{-i\omega_{mj} t_j} \sqrt{\frac{\gamma_{mj,mj}}{2}}\sigma_-^{mj}.
\end{align}
In the same way, we find
\begin{align}
a_{\text{out}}^\mathrm{L}(t)&=a_{\text{in}}^\mathrm{L}(t)+\sum_{mj}  e^{+i\omega_{mj} t_j} \sqrt{\frac{\gamma_{mj,mj}}{2}}\sigma_-^{mj}
\end{align}
for the output field propagating to the left. 
\section{Relaxation diagonalization}
\label{sec:relaxationDiagonalization}
In Sec.~\ref{sec:DressedBasis} of the main text we have shown that for a pair of qubits the master equation in its Lindblad form is
\be\begin{split}
\dot{\rho}
=&\frac{-i}{\hbar}\left[H,\rho\right]+\sum_{j,k=0,1}\gamma_{jk}\left[\sigma_-^{j}\rho \sigma_+^{k}-  \frac{1}{2}\left\{ \sigma_+^{k} \sigma_-^{j},\rho \right\}\right]\\
\equiv&\frac{-i}{\hbar}\left[H,\rho\right] + \mathcal{L}_\gamma\rho,
\end{split}\ee
where $H$ is given in Eq.~\eqref{eq:Heff:22}. In this appendix, we diagonalize the dissipator $\mathcal{L}_\gamma$ in order to find the dressed basis.

Diagonalization is achieved by using the standard approach of expressing the density matrix $\rho$ as a column vector in which case the dissipator takes the form
\be
\mathcal{L}_\gamma\vec{\rho}
= 
\sum_{kj}\gamma_{k,j}\left[\sigma_{b-}^{k} \sigma_{f+}^{Tj}  -\frac{1}{2}\sigma_{b+}^{j} \sigma_{b-}^{k}  -\frac{1}{2}\sigma_{f+}^{Tj} \sigma_{f-}^{Tk}\right]\vec{\rho},
\ee
with
\begin{align}
A \rho \rightarrow 1 \otimes A  \vec{\rho} = A_b \vec{\rho},\\ \nonumber
\rho A \rightarrow  A^T \otimes 1  \vec{\rho}=A_f^T \vec{\rho} \nonumber,
\end{align}
where $T$ refers to matrix transposition. In this way, the dissipator can be expressed as
\be
\mathcal{L}_\gamma\vec{\rho} 
=
\left[\sigma_{b-} \Upsilon \sigma_{f+}^T -\frac{1}{2}\sigma_{b+} \Upsilon \sigma_{b-}^T  -\frac{1}{2}\sigma_{f+} \Upsilon \sigma_{f-}^T \right]\vec{\rho},
\ee
with the Hermitian relaxation rate matrix
\be
\Upsilon= %
\begin{pmatrix}
  \gamma_{00} & \gamma_{01}\\ 
  \gamma_{01}^* & \gamma_{11}
 \end{pmatrix},
\ee
and where we have defined
\be
\sigma_{b\pm}= %
\begin{pmatrix}
  \sigma_{b\pm}^0 &  \sigma_{b\pm}^1
 \end{pmatrix},
\quad
\sigma_{f\pm}= %
\begin{pmatrix}
  \sigma_{f\pm}^{T0} &  \sigma_{f\pm}^{T1}
 \end{pmatrix}.
\ee
After diagonalizing $\Upsilon$ and going back to matrix form of the density matrix we find
\be
\mathcal{L}_\gamma\rho  =\sum_{i=B,D} \Gamma_{i} \mathcal D\left[ \sigma_-^{i} \right] \rho,
\ee
with the rates
\be
\Gamma_{B/D}=\frac{\gamma_{00}+\gamma_{11}}{2}\pm \sqrt{\left(\frac{\gamma_{00}-\gamma_{11}}{2}\right)^2+|\gamma_{01}|^2},
\ee
and dressed operators
\be
\sigma_-^{\mu}=\frac{\left(\Gamma_{\mu}-\gamma_{11} \right) \sigma_-^0 + \gamma_{01}^*\sigma_-^1}{\sqrt{\left(\Gamma_{\mu}-\gamma_{11} \right)^2+|\gamma_{01}|^2}},
\ee
for $\mu=B,D$.

\bibliographystyle{mybibtexstyle.bst} %Utiliser "nature" pour la version anglaise
\bibliography{referenceTransmissionLine.bib}

\end{document}